\documentclass[a4paper,11pt]{article}
\usepackage[utf8]{inputenc}
\usepackage[english]{babel}
\usepackage{amsmath}
\usepackage{amssymb}
\usepackage{amsfonts}
\usepackage{amsthm}
\usepackage{mathbbol}
\usepackage{graphicx}
\usepackage[dvipsnames]{xcolor}
\usepackage[final]{pdfpages} 
\usepackage[Symbol]{upgreek}
\numberwithin{equation}{section}

\DeclareSymbolFontAlphabet{\amsmathbb}{AMSb}

\allowdisplaybreaks                    
\def\tx{{\tt x}}
\def\ty{{\tt y}}

\def\hx{{\hat x}}
\def\hy{{\hat y}}
\def\hp{{\hat p}}
\def\heta{{\hat \eta}}

\def\kbar{{\mathchar'26\mkern-9muk}}

\def\tx{{\tt x}}                        
\def\bra#1{\langle #1 \vert}

\def\ket#1{\vert #1 \rangle}            
\def\p{\partial}                        
\def\pprime{{\prime\prime}}                

\def\dfrac #1#2{\displaystyle{\frac{#1}{#2}}}

\title{Quantum Scalar Field on Fuzzy de Sitter Space I\\
Field Modes and Vacua}

\author{Bojana Brki\'c $^{1}$, Ilija Buri\'c $^{2}$, Maja Buri\'c $^{1}$\
                           \ and
        Du\v sko Latas $^{1}$\thanks{bojana.brkic@ff.bg.ac.rs, ilija.buric@df.unipi.it, majab@ipb.ac.rs, latas@ff.bg.ac.rs}\
        \\[15pt]
        $\strut^{1}${\it Faculty of Physics, University of Belgrade, Studentski trg 12, 11001 Belgrade, Serbia}
        \\[5pt]
        $\strut^{2}${\it Department of Physics, University of Pisa, Largo B. Pontecorvo, 56127 Pisa, Italy}
        }

\begin{document}

\maketitle

\begin{abstract}
    We study a scalar field on a noncommutative model of spacetime, the fuzzy de~Sitter space, which is based on the algebra of the de~Sitter group $SO(1,d)$ and its unitary irreducible representations. We solve the Klein-Gordon equation in $d=2,4$ and show, using a specific choice of coordinates and operator ordering, that all commutative field modes can be promoted to solutions of the fuzzy Klein-Gordon equation. To explore completeness of this set of modes, we specify a Hilbert space representation and study the matrix elements (integral kernels) of a scalar field: in this way the complete set of solutions of the fuzzy Klein-Gordon equation is found. The space of noncommutative solutions has more degrees of freedom than the commutative one, whenever spacetime dimension is $d>2$. In four dimensions, the new non-geometric, internal modes are parametrised by $S^2\times W$, where $W$ is a discrete matrix space. Our results pave the way to analysis of quantum field theory on the fuzzy de~Sitter space.
\end{abstract}

\tableofcontents

\section{Introduction}
\label{S:Introduction}

The amount of astrophysical data describing the structure and the evolution of the universe is rapidly growing, both in quantity and precision, \cite{Planck:2018vyg}. In parallel, new theoretical methods are being developed, aiming in particular at a more precise description of the early inflationary expansion, \cite{Chen:2010xka,Cheung:2007st}. Since the correlations measured in the cosmic microwave background, as well as the distribution of the large scale structures, can be described by dynamics of a weakly coupled quantum scalar field, one of the main theoretical objectives of the moment is to calculate the two- and higher-point correlation functions at the end of inflation. These computations become more accessible, while in good agreement with observations, if the de~Sitter symmetry of spacetime is assumed. Much progress has been made in perturbative computations on  de Sitter background geometry, both in the in-in formalism and in the wavefunction approach. Furthermore, when viewed as conformal symmetry of the late-time boundary, de Sitter symmetry allows for a ‘bootstrap' approach to cosmological correlators, \cite{Baumann:2022jpr}.

Potentially important effects in the early universe cosmology might come from the interaction of scalar fields with gravity and the quantum nature of the latter. In the absence of the theory of quantum gravity, one may try to approximate quantised gravitational field by an effective or quantum spacetime. 

Imprints of spacetime noncommutativity on cosmological observations have been explored within several frameworks. In \cite{Chu:2000ww,Lizzi:2002ib} it was shown that Moyal-like deformations of coordinate commutators lead to inhomogeneities in the CMB, while \cite{Alexander:2001dr} observed that modified dispersion relations can lead to inflation without the need for the inflaton field. Effects of Moyal noncommutativity between minisuperspace variables rather than spacetime coordinates were studied in \cite{Garcia-Compean:2001jxk,Barbosa:2004kp}. The authors find the wave function of the universe by solving the deformed Wheeler-de Witt equation, and observe qualitative differences compared to commutative case.

A more fundamental approach, aiming in particular at resolution of cosmological singularity, is to realise cosmological spacetimes as large-$N$ limits of matrix models. Perturbations around these solutions give rise to noncommutative field theories on cosmological backgrounds. IKKT-inspired matrix models of two-dimensional and four-dimensional cosmologies were discussed in \cite{Chaney:2015ktw}: using rotationally-invariant ansatz, several noncommutative solutions and their commutative limits were found. A similar two-dimensional matrix cosmological model was proposed in \cite{Karczmarek:2022ejn}, where, along with geometry, modes of the scalar field, their propagation and quantisation were analysed. Implications of the IKKT model to cosmological perturbations and to gravitational waves were also discussed in \cite{Brahma:2021tkh}. 

Results obtained in the spectral action approach to noncommutative geometry were summarised in \cite{Marcolli:2018uea}: properties of various cosmological models, derived mainly from  ‘almost-commutative geometries' that are locally products of a commutative spacetime and a finite matrix geometry, are discussed.

Our aim in the present and the following papers is to study the effects on cosmological observables implied by modelling inflation as quantum field theory on fuzzy de~Sitter space, introduced in \cite{Buric:2017yes}. The reasoning, broadly speaking, is to give to spacetime a kind of discrete structure by representing coordinates by non-commuting operators, while preserving classical symmetries. Technically, this is done by using the generators of the de~Sitter group and its irreducible representations, to define the fuzzy de~Sitter space. Among several versions of noncommutative geometry, the fuzzy de~Sitter space is constructed in the framework of noncommutative frame formalism \cite{Madore:2000aq}. This formalism gives the basic differential and Riemannian geometry of a noncommutative space (differential forms, connection, curvature, Laplace operator), as well as a description of classical scalar and gauge fields on it. Thus, our approach may be viewed as being in between those from matrix models and the ones based on Moyal deformations. In contrast to the former, we do not commit to particular microscopic physics that may give rise to the noncommutative space, thus keeping the analysis more general. Nevertheless, it would certainly be worthwhile to investigate whether the fuzzy de~Sitter space arises from a matrix model. The main difference compared to Moyal deformation approaches lies in the $SO(1,4)$ symmetry of our spacetime, which allows to study it using powerful representation-theoretic methods and eventually bootstrap techniques, akin to \cite{Hogervorst:2021uvp,DiPietro:2021sjt}.

Noncommutative geometry of fuzzy de~Sitter space and quantum mechanics on it have been discussed previously in \cite{Buric:2017yes,Brkic:2021lre}, while some cosmological implications were considered in \cite{Buric:2015wta,Buric:2019yau}. Our goal is to develop the theory of a quantum scalar field on this background and compute late-time correlation functions. These can then be compared to correlators on classical de~Sitter space and eventually to observations. In the present work, we shall make the first step in this direction by 1) obtaining the complete set of free field modes, i.e. eigenfunctions of the fuzzy Laplace operator and 2) defining a corresponding $SO(1,4)$-invariant vacuum. These results open the way to computing correlation functions -- the simplest example, that of a two-point function, is discussed in the concluding section. In the context of noncommutative geometry, our model provides a four-dimensional noncommutative space with a covariant differential calculus in which various exact calculations are possible. 

Let us describe the content of the paper in more details. Section \ref{S:Coordinates on dSd field quantisation and vacua} is concerned with the  commutative de~Sitter space. Our main objective is to discuss field modes, i.e. solutions to the Klein-Gordon equation, in a way that generalises to the noncommutative case. For this purpose we shall introduce a set of coordinates $(\upeta,\ty^i)$, for which no difficulties due to operator ordering arise
in the process of quantisation. We shall write down an orthonormal basis of solutions to the Klein-Gordon equation separated in $(\upeta,\ty^i)$ coordinates and compare these to solutions in the usual  Poincar\'e coordinates by computing the corresponding Bogoliubov coefficients. It is shown that the natural choice of vacuum in the $(\upeta,\ty^i)$ coordinates is invariant under the de~Sitter group; a different choice of positive frequency modes that gives rise to the Bunch-Davies vacuum is identified. This section essentially does not depend on the spacetime dimension and the latter is kept general.

The remainder of the paper carries out the same analysis in the noncommutative setup. As mentioned, we work in the frame formalism, focusing on the fuzzy dS$_4$ space of \cite{Buric:2017yes}. We shall also study the simpler case of the two-dimensional fuzzy dS$_2$, whose definition is strictly analogous to the $h$-deformed hyperbolic plane, \cite{cho,Madore:1999fi}. On both spaces, coordinates are defined as operators in a certain irreducible representation of the classical isometry group $SO(1,d)$. Differential geometry is derived from a set of  momenta, elements of $\mathfrak{so}(1,d)$, and closely resembles that of the commutative space: for example, we may say that fuzzy de Sitter spaces satisfy Einstein's equations with a positive cosmological constant. Elements of the geometry necessary for the study of harmonic functions, in particular the definition of the Laplace-Beltrami operator, are reviewed at the beginning of Section \ref{S:Fuzzy harmonics: solutions in noncommutative coordinates}. We then proceed to show that eigenfunctions of the Laplacian on the commutative dS$_d$, separated in the $(\upeta,\ty^i)$ coordinates, may be directly quantised to give eigenfunctions (eigenoperators) of the fuzzy Laplacian. The idea is to use commutation relations between coordinates and momenta to prove that classical solutions do not suffer from operator ordering issues and remain Laplacian eigenfunctions on the quantum level. A favourable property of thus obtained fuzzy harmonics is their direct link to commutative ones, the latter being recovered from the former in the commutative limit.

The fuzzy Klein-Gordon equation is an equation for functions whose arguments are operators. Although the similarity of its structure with the classical Klein-Gordon equation gives a way to quantise classical solutions, when solved purely algebraically it cannot answer the question about the  completeness of the set of solutions. To answer this question, in Section 4 we present another study of fuzzy harmonics, based on an explicit realisation of the fuzzy space by operators on a Hilbert space. In this representation, operators and functions thereof are written as integral kernels or infinite-dimensional matrices, and the fuzzy Klein-Gordon equation becomes a partial differential equation. A simple counting reveals that the Laplacian on fuzzy dS$_4$ possesses more eigenfunctions than its commutative counterpart. Furthermore, Klein-Gordon equations in the two cases take the same form -- the only difference is that fuzzy harmonics come multiplied by an arbitrary function on a two-sphere, with further discrete degrees of freedom. In particular, the set of eigenfunctions found in Section \ref{S:Fuzzy harmonics: solutions in noncommutative coordinates} is not complete: a complete basis of solutions is given in Section \ref{S:All fuzzy harmonics}. This mismatch in the number of modes is a manifestation of the fact that the fuzzy dS$_4$ has more `internal' degrees of freedom than dS$_4$, a commonplace for higher-than-two dimensional noncommutative spaces. The fact that the internal space is compact suggests organising the noncommutative degrees of freedom as a discrete tower of KK modes over the commutative ones.

The drawback of the integral kernel method is that the commutative limit is obscured. Section \ref{S:All fuzzy harmonics} goes some way towards showing how the two approaches are related -- to this end, we compute integral kernels corresponding to solutions of Section \ref{S:Fuzzy harmonics: solutions in noncommutative coordinates}. More precisely, all integral kernels are computed in the two-dimensional case, while in four dimensions we restrict to a subset thereof. 

Results of this work mostly clear the path towards computations of correlation functions in quantum field theory: for instance, two-point functions can be written as integrals over field modes, and are thus readily accessible. That is, in principle, up to mathematical complexities that arise in the  realistic four-dimensional model: to systematically approach these computations, we have to develop specific methods, presumably using group and representation theory. Another point that requires  further study is the role of internal modes, as well as the commutative limit of fuzzy observables, e.g. propagator. The semi-classical states and the fuzzy propagator are discussed in the two-dimensional case in the concluding Section \ref{S:Summary and perspectives}, where we also summarise the obtained results and outline plans for the future.

\section{Coordinates on dS$_d$, field quantisation and vacua}
\label{S:Coordinates on dSd field quantisation and vacua}

In this section we shall study the Klein-Gordon equation on the $d$-dimensional de~Sitter space: the novelty of our analysis lies in the use of a set of coordinates that are suggested by noncommutative geometry. In the first subsection, we introduce the usual ‘Poincar\'e' coordinates $(\upeta,\tx^i)$ and the new $(\upeta,\ty^i)$ coordinates, and write the Laplacian in the two coordinate systems. We solve the Klein-Gordon equation and compute overlaps between field modes in $(\upeta,\tx^i)$ and $(\upeta,\ty^i)$ coordinates. In both coordinate systems, the space of solutions is naturally divided into two orthogonal $SO(1,d)$-invariant subspaces that carry (isomorphic) unitary irreducible representations of the de~Sitter group. The vacua that correspond to these solutions are therefore de~Sitter invariant.

{\bf Remark\,} Throughout the text we will be concerned with several quantities of similar physical meaning: these are distinguished by using various fonts. For example, $(\upeta,\ty^i)$ denote coordinates on the commutative de~Sitter space dS$_d$; $(\heta,\hy^i)$ are their noncommutative counterparts, operators on the fuzzy de~Sitter space (a noncommutative algebra $\cal{A}$), while $(\xi, y^i)$ are certain variables used for constructing a representation space $\cal{H}$ of $\cal{A}$.

\subsection{Coordinate systems}

We consider the $d$-dimensional de Sitter space. More precisely, we shall only work with one half of the space, which can be parametrised by conformally flat coordinates,
\begin{equation}\label{Poincare-coords}
    ds^2 = \frac{\alpha^2}{\upeta^2}\, \left(-d\upeta^2 + d\tx^i d\tx_i\right)\,, \qquad \upeta\in(-\infty,0),\ \ \tx^i\in(-\infty,\infty)\ .
\end{equation}
In the context of cosmology, this space models the inflationary phase of the universe and the asymptotic region $\{\eta = 0\}$ is the reheating surface. We will refer to $(\upeta,\tx^i)$ as Poincar\'e coordinates. Indices $i,j=1,\dots,d-1\,$ are raised and lowered with the flat Euclidean metric. In the following, we will also make use of another set of coordinates $(\upeta,\ty^i)$, related to the above by
\begin{equation}\label{eta-y-coords}
   \ty^i =\frac{\tx^i}{\upeta}\,, \qquad ds^2 =\frac{\alpha^2}{\upeta^2}\,\Big(-\big(1-\ty^2\big)\, d\upeta^2 +2\upeta\, \ty^i\, d\upeta d\ty_i + \upeta^2 d\ty^i d\ty_i \Big) \ .
\end{equation}
Throughout this work we will fix the cosmological constant by setting $\alpha=1$; only occasionally we reinstate $\alpha$-dependence in certain equations to increase physical clarity. The isometry group of the de Sitter space is $G = SO(1,d)$. This group acts on the asymptotic boundary by conformal transformations and we shall often denote its generators  in conformal-like notation, $\{P_i,K_i,L_{ij},D\}$. Relation between these generators and Lorentz-like ones, as well as other conventions regarding the group $SO(1,d)$, are collected in Appendix \ref{A:Conventions for the de Sitter algebra}. The Lie algebra $\mathfrak{g} = \text{Lie}(G)$ acts on dS$_d$ by vector fields. In Poincar\'e coordinates these vector fields are
\begin{align}\label{isometries-dSd-1}
    & P_i = - \partial_{\tx^i}, \quad L_{ij} = \tx_i \partial_{\tx^j} - \tx_j \partial_{\tx^i}\,,\\[2pt]
    & D = -\upeta\partial_\upeta -\tx^i \partial_{\tx^i}, \quad K_i = (\upeta^2 - \tx^2) \partial_{\tx^i} - 2\tx_i D\,,\label{isometries-dSd-2}\\
\intertext{while in coordinates $(\upeta,\ty^i)$, the action reads}
    & P_i = -\frac{1}{\upeta}\,  \partial_{\ty^i}, \quad L_{ij} = \ty_i\partial_{\ty^j} - \ty_j\partial_{\ty^i}\,,\\
    & D = -\upeta\partial_\upeta, \quad K_i = \upeta \left(1-\ty^2\right)\partial_{\ty^i} + 2\upeta^2 \ty_i \partial_\upeta\ .
\end{align}

Quantum field theory on the de Sitter background requires the knowledge of free field modes, i.e. eigenfunctions of the Laplace-Beltrami operator. For this reason, we spell out the Laplacian in the two coordinate systems introduced above. The operator is found from the usual formula
\begin{equation}\label{Laplacian-general}
    \Delta_g = \frac{1}{\sqrt{|g|}} \partial_\mu \sqrt{|g|} g^{\mu\nu} \partial_\nu\ .
\end{equation}
In Poincar\'e coordinates, this gives
\begin{equation}\label{Laplacian-Poincare}
    \Delta_{dS_d} = -\upeta^2\partial_\upeta^2 + \upeta^2 \partial_{\tx^i}\partial_{\tx_i} + (d-2)
\upeta\partial_\upeta\,,
\end{equation}
while in $(\upeta,\ty^i)$ coordinates, the Laplacian reads
\begin{equation}\label{Laplacian-eta-y}
    \Delta_{dS_d} =  - \ty^i \ty^j \partial_{\ty^i} \partial_{\ty^j} + \partial_{\ty^i}\partial_{\ty_i}  - \upeta^2 \partial_\upeta^2 + 2\upeta \partial_\upeta\ty^i \partial_{\ty^i}- d \,\ty^i \partial_{\ty^i} + (d-2)\upeta\partial_\upeta\ .
\end{equation}
The Laplacian commutes with action of $SO(1,d)$ generators and in fact coincides with the quadratic Casimir constructed out of them, see e.g. \cite{Kirillov}.

\subsection{Review: quantisation of scalar field and Bogoliubov coefficients}

We consider a theory of a single real scalar field $\Phi$ of mass $M$ on the dS$_d$ background. The starting point for such an analysis is the space of complex solutions to the Klein-Gordon equation
\begin{equation}\label{KG-equation}
    \left(\Delta_{dS_d} - M^2\right) \Phi = 0\ .
\end{equation}
This space carries the bilinear form (by the common abuse of terminology, we shall refer to it as the Klein-Gordon inner product), defined via integration over a Cauchy surface $\Sigma$. If we choose the surface to be a time slice $\Sigma=\{\upeta=\text{const}\}$, then
\begin{equation}\label{KG-inner-product}
    \langle \Phi_1,\Phi_2\rangle = -i \int_\Sigma d^{d-1}\tx\ \upeta^{-d+2}\left(\Phi_1^\ast \partial_\upeta \Phi_2 - \Phi_2 \partial_\upeta \Phi_1^\ast\right)\ .
\end{equation}
Field equations \eqref{KG-equation} imply that the form \eqref{KG-inner-product} is independent of the choice of the spacelike surface $\Sigma$. In de~Sitter space, it is convenient to use this freedom to move $\Sigma$ to the asymptotic future $\upeta\to0$ in  Poincar\'e coordinates. Behaviour of solutions in this limit is easily found from the asymptotics of the differential operator \eqref{Laplacian-Poincare},
\begin{equation}
    \Delta_{dS_d} \sim  -\upeta^2\partial_\upeta^2 + (d-2)\upeta\partial_\upeta\ .
\end{equation}
Every solution can be written in the form
\begin{equation}\label{asymptotics-of-solutions}
    \Phi(\upeta,\tx^i) \sim (-\upeta)^{\mathbb{\Delta}_+}\Phi^+(\tx^i) + (-\upeta)^{\mathbb{\Delta}_-}\Phi^-(\tx^i)\,, \qquad \upeta\to 0\,,
\end{equation}
where
\begin{equation}
    \mathbb{\Delta}_\pm = \frac{d-1}{2} \pm i\kappa\,, \qquad \kappa^2 =M^2 - \left(\frac{d-1}{2}\right)^2 \,,
\end{equation}
(see e.g. \cite{Sleight:2019hfp}). Conversely, given a pair of functions $\Phi^\pm(\tx^i)$, there is a solution to the Klein-Gordon equation with the asymptotic behaviour \eqref{asymptotics-of-solutions}. Therefore, we may label solutions by such pairs of functions. We shall denote
\begin{equation}\label{two-component-notation}
    \Phi(\upeta,\tx^i) \cong \begin{pmatrix} 
    \Phi^+(\tx^i)\\[4pt]
    \Phi^-(\tx^i)
    \end{pmatrix}\,,
\end{equation}
to designate this relation and often refer to pairs $\,(\Phi^+(\tx^i),\Phi^-(\tx^i))^T\,$ simply as ‘solutions'. In terms of the asymptotic functions $\Phi^\pm(\tx^i)$, the scalar product \eqref{KG-inner-product} becomes
\begin{equation}\label{boundary-scalar-product}
    \langle \Phi_1,\Phi_2\rangle = 2\kappa \int d^{d-1}\tx\ \Big((\Phi_1^+)^\ast \Phi_2^+ - (\Phi_1^-)^\ast\Phi_2^-\Big)\ .
\end{equation}
As the Laplacian commutes with the action of isometries of dS$_d$, the space of solutions to \eqref{KG-equation} carries a representation of $SO(1,d)$. The representation consists of two irreducible components, isomorphic to one another, which belong to the unitary principal series, \cite{Dobrev:1977qv}.\footnote{We assume that $\kappa^2>0$ i.e. that the scalar field is heavy, $M^2>(d-1)^2/4$, and accordingly we use the principal series representations of $SO(1,d)$. Whether $\kappa$ is real or imaginary is relevant for transformation properties of the Bessel functions which we will use; the end results are often vary similar for light and heavy fields.} This fact is manifested in the notation \eqref{two-component-notation} -- each of the component functions $\Phi^+(\tx^i)$ and $\Phi^-(\tx^i)$ parametrises a vector in one of the two principal series representations of $SO(1,d)$.\footnote{In the notation used in earlier papers \cite{Buric:2017yes,Brkic:2021lre}, quantum numbers of the representations in $d=4$ are $\rho=\kappa$, $s=0$.} From \eqref{boundary-scalar-product} it is also clear that the Klein-Gordon inner product \eqref{KG-inner-product} is $SO(1,d)$-invariant.

Quantisation of the scalar field proceeds in the usual way, \cite{Birrell:1982ix,Wald:1995yp}. We expand the field in modes and promote coefficients to creation and annihilation operators acting on a Fock space,
\begin{equation}\label{mode-expansion-1}
    \Phi(x) = \sum_i \Big(a_i u_i(x) + a_i^\dagger u_i^\ast(x) \Big) \, , \qquad  a_i\ket{0}_a=0\ .
\end{equation}
The index $a$ on the vacuum indicates that it is annihilated by the set of operators $\{a_i\}$. The modes are arbitrary up to the requirement
\begin{equation}\label{orthogonality-modes}
    \langle u_i, u_j \rangle = \delta_{ij}, \quad \langle u_i^\ast, u_j^\ast \rangle = -\delta_{ij}, \quad \langle u_i,u_j^\ast \rangle = 0\ .
\end{equation}
Bogoliubov coefficients arise upon quantising with respect to two different sets of modes. Let
\begin{equation}\label{mode-expansion-2}  
    \Phi(x) = \sum_i \Big( b_i v_i(x) +b_i^\dagger v_i^\ast(x) \Big)\, , \qquad  b_i\ket{0}_b=0\,,
\end{equation}
be a second expansion satisfying orthogonality relations analogous to \eqref{orthogonality-modes}. Bogoliubov coefficients $\alpha_{ji},\beta_{ji}$ are the overlaps between the sets $\{u_i,u_i^\ast\}$ and $\{v_i,v_i^\ast\}$,
\begin{align}
    & v_j = \sum_i \Big(\alpha_{ji} \,u_i +\beta_{ji} \,u_i^\ast\Big) \,, \qquad  a_i = \sum_j  \Big(\alpha_{ji} \,b_j +\beta_{ji} \,b_j^\dagger\Big)\,, \label{Bogoliubov}\\
    & u_i = \sum_j \Big(\alpha_{ji}^\ast\, v_j - \beta_{ji}\, v_j^\ast\Big) \,, \qquad b_j = \sum_i  \Big(\alpha_{ji}^*\, a_i - \beta_{ji}^*\, a_i^\dagger\Big)\ .\label{Bogoliubov-inv}
\end{align}
In particular, if coefficients $\beta_{ji}$ are non-vanishing, the vacuua $|0\rangle_a$, $|0\rangle_b$ of the two quantisations are different. The number of $a$-particles in the $b$-vacuum is
\begin{equation}\label{broj-ces}
    \bra{0}_b\, N_{i,a}\,\ket{0}_b = \bra{0}_b\,\, a_i^\dagger a_i\,\ket{0}_b=\sum_j\vert \beta_{ji} \vert ^2\ .           
\end{equation}

\subsection{Field modes in Poincar\'e coordinates, the Bunch-Davies vacuum}

We turn to solutions to the Klein-Gordon equation, working first in  Poincar\'e coordinates. Let $\,h_{k,\lambda}(\tx^i)$ be an eigenfunction of the flat-space Laplacian $\partial_{\tx^i}\partial_{\tx_i}$ with the eigenvalue $-k^2$. Labels $\lambda$ are additional quantum numbers that such functions carry. For example, in spherical polar coordinates
\begin{equation}\label{h-ipsilon}
    h_{k,l,\vec{m}}(\tx^i) = \sqrt{k}\, r^{\frac{3-d}{2}} J_{\frac{d-3}{2}+l}(kr) \, Y_l^{\vec{m}}(\theta_a)\,,
\end{equation}
where $Y_l^{\vec{m}}(\theta_a)$ are spherical harmonics on the sphere $S^{d-2}$. The Klein-Gordon equation is solved by
\begin{equation}
    u(\upeta,\tx^i) = (-\upeta)^{\frac{d-1}{2}} \big(C_+  J_{i\kappa}(-k\upeta) + C_- J_{-i\kappa}(-k\upeta)\big) \, h_{k,\lambda}(\tx^i)\,,
\end{equation}
where $J_{\pm i\kappa}$ are Bessel functions. We shall denote
\begin{equation}\label{modes-u}
    u_{k,\lambda,\pm\kappa}(\upeta,\tx^i) = C_\pm (-\upeta)^{\frac{d-1}{2}} J_{\pm i\kappa}(-k\upeta) \, h_{k,\lambda}(\tx^i)\ .
\end{equation}
The asymptotic behaviour of functions $u_{k,\lambda,\pm\kappa}(\upeta,\tx^i) $ in the limit $\upeta\to 0$ is determined by properties of the Bessel functions,
\begin{equation}
    (-\upeta)^{\frac{d-1}{2}} J_{\pm i\kappa}(-k\upeta) \sim (-\upeta)^{\frac{d-1}{2} \pm i\kappa} \, \frac{(k/2)^{\pm i\kappa}}{\Gamma(1\pm i\kappa)}\ .
\end{equation}
We see that field modes in the boundary limit greatly simplify. This will be a recurring theme until the end of the section. Using the notation introduced in \eqref{two-component-notation}, we can write
\begin{equation}\label{boundary-solutions-Poincare}
    u_{k,\lambda,\kappa} \cong \frac{C_+ (k/2)^{i\kappa}}{\Gamma(1+i\kappa)} \begin{pmatrix}  
    h_{k,\lambda}(\tx^i)\\
    0
    \end{pmatrix}, \quad  u_{k,\lambda,-\kappa} \cong \frac{C_- (k/2)^{-i\kappa}}{\Gamma(1-i\kappa)} \begin{pmatrix} 
    0\\
     h_{k,\lambda}(\tx^i)
     \end{pmatrix} \ .
\end{equation}
Clearly, functions with the index $\kappa$ are orthogonal to those with the index $-\kappa$. Furthermore, functions in each of the sets are orthogonal, with the normalisation
\begin{equation}\label{inner-products-intermediate}
    \langle u_{k,\lambda,\kappa},u_{k',\lambda',\kappa} \rangle = -\langle u_{k,\lambda,-\kappa},u_{k',\lambda',-\kappa} \rangle = \vert C_\pm\vert ^2\, \frac{2\sinh(\pi\kappa)}{\pi} \, \delta(k-k') \, \delta_{\lambda\lambda'} \ .
\end{equation}
We set $\, C_+=C_-= \sqrt{\pi/(2\sinh(\pi\kappa))}$. Spaces $\{u_{k,\lambda,\kappa}\}$ and $\{u_{k,\lambda,-\kappa}\}$ form mutually orthogonal subspaces of `positive' and `negative' frequency solutions to \eqref{KG-equation}. As $\, J^\ast_{\pm i\kappa}(-k\upeta) =J_{\mp i\kappa}(-k\upeta)$, complex conjugation maps these two spaces to one another, which means that the above modes are appropriate for quantisation, \cite{Wald:1995yp}. To be precise, the modes satisfy $\,u^\ast_{k,\lambda,\kappa} = u_{k,\lambda,-\kappa}\, $ only if the spatial parts of 
 (\ref{h-ipsilon}) are real,
\begin{equation}\label{psi-h}
    h^\ast_{k,\lambda}(\tx^i)= h_{k,\lambda}(\tx^i)\ .
\end{equation}
One can always choose a set of real orthonormal modes $ h_{k,\lambda}$. To conform with the standard notation, \cite{Birrell:1982ix}, we will assume that the this is indeed the case. In $d=4$ the reality requirement amounts to using, instead of spherical harmonics $Y_l^m$, their linear combinations $\Psi_l^m$ defined by
\begin{equation}
    \Psi_l^m(\theta,\varphi) = \frac{e^{\frac{i\pi m}{2}} }{\sqrt 2}\, \left(e^{\frac{i\pi}{4}}Y_l^m(\theta,\varphi) + e^{-\frac{i\pi}{4}}Y_l^{-m}(\theta,\varphi)\right) \ .
\end{equation}
Manifestly from \eqref{boundary-solutions-Poincare}, the set of positive frequency modes, and thereby the corresponding vacuum, is $SO(1,d)$-invariant. 

The common choice of  positive frequency solutions in cosmology is defined by the behaviour of modes at large momenta, $|k\eta| \to \infty$. It is the Bunch-Davies vacuum,
\begin{align}\label{BD-solutions-1}
    & u_{k,\lambda,\kappa}^{BD} = \frac{\sqrt{\pi}}{2}e^{-\frac{\pi\kappa}{2}}(-\upeta)^{\frac{d-1}{2}} H^{(1)}_{i\kappa}(-k\upeta)\, h_{k,\lambda}(\tx^i) = \frac{1}{\sqrt{2\sinh(\pi\kappa)}} \left( e^{\frac{\pi\kappa}{2}}u_{k,\lambda,\kappa} - e^{-\frac{\pi\kappa}{2}} u_{k,\lambda,-\kappa}\right)\,,\\
    & (u_{k,\lambda,\kappa}^{BD})^\ast = \frac{\sqrt{\pi}}{2}e^{-\frac{\pi\kappa}{2}} (-\upeta)^{\frac{d-1}{2}} H^{(2)}_{-i\kappa}(-k\upeta) \, h_{k,\lambda}^\ast(\tx^i) = \frac{1}{\sqrt{2\sinh(\pi\kappa)}} \left( e^{\frac{\pi\kappa}{2}} u_{k,\lambda,\kappa}^\ast - e^{-\frac{\pi\kappa}{2}} u_{k,\lambda,-\kappa}^\ast\right)\ .\label{BD-solutions-2}
\end{align}
The inner product \eqref{inner-products-intermediate} implies that the Bunch-Davies modes are orthonormal,
\begin{equation}
    \langle u_{k,\lambda,\kappa}^{BD},u_{k',\lambda',\kappa}^{BD} \rangle  = \delta(k-k')\, \delta_{\lambda \lambda'}\,, \qquad \langle u_{k,\lambda,\kappa}^{BD},(u_{k',\lambda',\kappa}^{BD})^\ast \rangle = 0\ ,
\end{equation}
in accord with the requirements \eqref{orthogonality-modes}. The Bunch-Davies vacuum is $SO(1,d)$-invariant as well. This can be seen e.g. from the expansion of the $u_{k,\lambda,\kappa}$ in the Bunch-Davies modes. If we define the corresponding Bogoliubov coefficients as
\begin{equation}\label{alpha-transformation}
      u_{k,\lambda,\kappa} = \sum_{k'\lambda'} \Big(\alpha_{k\lambda,k'\lambda'} \,  u_{k',\lambda',\kappa}^{BD} +\beta_{k\lambda,k'\lambda'} \,  ( u_{k',\lambda',\kappa}^{BD})^\ast\Big) \ , 
\end{equation}
we find that their values are constant, independent of quantum numbers $k$ and $\lambda$,
\begin{equation}\label{u-BD}
    \alpha_{k\lambda,k'\lambda'}=\frac {e^{\frac{\pi\kappa}{2}}}{\sqrt{2\sinh(\pi\kappa)}}\, \delta(k-k')\, \delta_{\lambda\lambda'}\, ,
    \quad    \beta_{k\lambda,k'\lambda'}=\frac {e^{-\frac{\pi\kappa}{2}}}{\sqrt{2\sinh(\pi\kappa)}}\, \delta(k-k')\, \delta_{\lambda\lambda'} \ .
\end{equation}
Since the two representations of $SO(1,d)$ spanned by $\{u_{k,\lambda,\kappa}\}$ and $\{u^\ast_{k,\lambda,\kappa}\}$ are isomorphic to each other, the constant linear combinations $\,u^{BD}_{k,\lambda,\kappa}$ of these modes again form a representation of $SO(1,d)$ and the corresponding vacuum is de~Sitter invariant. Stated in another way, the vacuum defined by $u_{k,\lambda,\kappa}$ belongs to the family of `$\alpha$-vacua' of \cite{Allen:1985ux} (see also \cite{Chernikov:1968zm}) with
\begin{equation}\label{alpha}
    \coth\alpha = e^{\pi\kappa} \ .
\end{equation}
The corresponding Green's functions (propagator, Hadamard function etc.) are easily obtained from the Bunch-Davies two-point function, \cite{Allen:1985ux}.

\subsection{Field modes in coordinates $(\upeta,\ty^i)$}

We turn now to the Klein-Gordon equation in coordinates $(\upeta,\ty^i)$. In these coordinates, it is natural to look for solutions in the separated form,
\begin{equation}\label{Ansatz}
    v(\upeta,\ty^i) = (-\upeta)^{-i\omega} F(\rho) \, Y_l^{\vec{m}}(\theta_a)\,,
\end{equation}
where $(\rho,\theta_a)$ are spherical polar coordinates constructed from $\ty^i$ (in particular, $\rho=r/| \upeta |$). By construction, functions \eqref{Ansatz} diagonalise the ‘dilation operator' $D$, in addition to the Laplacian $\Delta_{dS_d}$. In two dimensions, this property, together with boundary conditions, completely determines modes $v(\upeta,\ty^i)$, whereas in four dimensions functions \eqref{Ansatz} may be characterised as simultaneous eigenfunctions of the complete set of commuting operators $\{\Delta_{dS_4},D,L_{ij}L^{ij},L_{12}\}$. The reduction of the Laplacian to the radial function $F(\rho)$ gives
\begin{equation}\label{KG-spherical}
    \Delta^{-i\omega,l,\vec{m}}_{dS_d} = \left(1 - \rho^2\right) \partial_\rho^2 + \Big(\frac{d-2}{\rho} - (d + 2i\omega)\rho\Big)\partial_\rho - \frac{l(l+d-3)}{\rho^2} -i\omega(d-1+i\omega)\ .
\end{equation}
Eigenfunctions of this operator are given in terms of the hypergeometric function $_2F_1$; the
solution finite at $\rho=0\,$ is
\begin{equation}\label{KG-sferne}
F(\rho) =\rho^l\,_2F_1\Big( \frac{2l +d-1+2i(\omega-\kappa)}{4}\, ,\,  \frac{2l+ d-1+2i(\omega+\kappa)}{4}\, ;\, \frac{d-1}{2}+l \, ;\rho^2\Big) \ .       
\end{equation}
However, since our solutions are arranged  according to their behaviour at late times $\upeta\to 0$, it is more convenient to use the variable $\,\rho^{-1}$. The general solution is of the form
\begin{align}
    & v (\upeta,\rho,\theta_a) = (-\upeta)^{-i\omega} \left(c_+\ \rho^{-\frac{d-1}{2}-i(\omega+\kappa)}\ _2F_1\Big(\frac{2l+d-1 + 2i(\omega+\kappa)}{4},\frac{-2l+5-d+2i(\omega+\kappa)}{4},1+i\kappa,\rho^{-2}\Big)\right.  \nonumber\\ 
    &\left. + \, c_-\ \rho^{-\frac{d-1}{2}-i(\omega-\kappa)}\ _2F_1\Big(\frac{2l+d-1 + 2i(\omega-\kappa)}{4},\frac{-2l+5-d+2i(\omega-\kappa)}{4},1-i\kappa,\rho^{-2}\Big)\right)Y_l^{\vec{m}}(\theta_a)\ .  \nonumber
\end{align}
Similarly as above, we introduce modes with definite asymptotics, 
\begin{align}\label{solutions-(eta,y)-coords}
    v_{\omega,l,\vec{m},\kappa} =\ & c\, (-\upeta)^{-i\omega}\, \rho^{-\frac{d-1}{2}-i(\omega+\kappa)}\nonumber\\
    & _2F_1\Big(\frac{2l+d-1 + 2i(\omega+\kappa)}{4},\frac{-2l+5-d+2i(\omega+\kappa)}{4},1+i\kappa,\rho^{-2}\Big) \Psi_l^{\vec{m}}(\theta_a) \ ,
\end{align}
where $\Psi_l^{\vec{m}}$ are real linear combinations of $Y_l^{\vec{m}} $ with fixed $l$. In the limit $\,\upeta\to 0\,$, keeping $r$ constant, $\rho^{-1}$ approaches zero, and the solution behaves as
\begin{equation}
    v_{\omega,l,\vec{m},\pm\kappa} \sim  c \, (-\upeta)^{\frac{d-1}{2} \pm i\kappa}\, r^{-\frac{d-1}{2}\mp i\kappa  -i\omega} \, \Psi_l^{\vec{m}}(\theta_a)\ .
\end{equation}
In the two-component notation \eqref{two-component-notation} we may write
\begin{equation}\label{boundary-solutions-Descartes}
    v_{\omega,l,\vec{m},\kappa} \cong \begin{pmatrix}
   c\, r^{-\frac{d-1}{2}- i\kappa -i\omega}\, \Psi_l^{\vec{m}}(\theta_a)\\
    0
    \end{pmatrix}, \quad v_{\omega,l,\vec{m},-\kappa}\cong \begin{pmatrix}
    0\\
    c \, r^{-\frac{d-1}{2}+ i\kappa -i\omega} \, \Psi_l^{\vec{m}}(\theta_a)
    \end{pmatrix}\ .
\end{equation}
Thus, using  \eqref{boundary-scalar-product} for the scalar product on the asymptotic boundary, we find
\begin{equation*}\label{orthogonality-relations-etay}
   \langle v_{\omega,l,\vec{m},\kappa}, v_{\omega',l',\vec{m}',\kappa} \rangle = - \langle v^\ast_{\omega,l,\vec{m},\kappa}, v^\ast_{\omega',l',\vec{m}',\kappa} \rangle = 4\pi\kappa \vert c\vert ^2\,\delta(\omega - \omega')\, \delta_{ll'}\delta_{\vec{m}\vec{m}'}\,, \quad \langle v_{\omega,l,\vec{m},\kappa}, v^\ast_{\omega',l',\vec{m}',\kappa} \rangle = 0\ .
\end{equation*}
Functions $\{v_{\omega,l,\vec{m},\kappa}\}$ form a valid set of positive frequency solutions; we set $\, c=1/\sqrt{4\pi\kappa}\,$. 

From the boundary expressions \eqref{boundary-solutions-Poincare} and \eqref{boundary-solutions-Descartes} it is clear that the positive frequency solutions $v_{\omega,l,\vec{m},\kappa}$ are linear combinations of the positive frequency solutions $u_{k,l,\vec{m},\kappa}$ -- both sets span the space of solutions whose second component in \eqref{two-component-notation} vanishes. This means that the vacua defined by the two sets are the same (and different from the Bunch-Davies vacuum). Thus, solutions to the Klein-Gordon equation in $(\upeta,\ty^i)$ coordinates naturally give a de~Sitter invariant vacuum. One can compute the Bunch-Davies modes in $(\upeta,\ty^i)$ coordinates by inversion of the $\alpha$-transformation \eqref{alpha-transformation}. This leads to
\begin{equation}
    v_{\omega,l,\vec{m},\kappa}^{BD} = \cosh\alpha\,  v_{\omega,l,\vec{m},\kappa}-\sinh \alpha \,v_{\omega,l,\vec{m},\kappa}^\ast = \frac {1}{\sqrt{2\sinh(\pi\kappa)}}\left(e^{\frac{\pi\kappa}{2}} v_{\omega,l,\vec{m},\kappa}-e^{\frac{-\pi\kappa}{2}}v_{\omega,l,\vec{m},\kappa}^\ast\right)\,,
\end{equation}
where, recall, $\alpha$ is given by \eqref{alpha}. We may go further and obtain the overlaps between $u$-modes and $v$-modes. It suffices to find the Bogoliubov coefficients
\begin{equation}
    \alpha_{\omega l\vec{m},kl'\vec{m}'} = \langle v_{\omega, l ,\vec{m},\kappa},u_{k,l',\vec{m}',\kappa}^{BD} \rangle, \qquad 
    \beta_{\omega l\vec{m},kl'\vec{m}'} = \langle v_{\omega, l, \vec{m},\kappa}, (u_{k,l',\vec{m}',\kappa}^{BD})^\ast \rangle\ .
\end{equation}
Overlaps involving mode functions $\{v_{\omega,l,\vec{m}}^{BD},(v_{\omega,l,\vec{m}}^{BD})^\ast\}$ and $\{u_{k,l,\vec{m},\kappa},u^\ast_{k,l,\vec{m},\kappa}\}$ are obtained from these with the help of \eqref{alpha-transformation} and \eqref{alpha}. The  coefficients are found to be
\begin{align}
    \alpha_{\omega l\vec{m},kl'\vec{m}'} & = \frac{\sqrt{\kappa}\,e^{\frac{\pi\kappa}{2}}}{2\sinh(\pi\kappa)\,\Gamma(1+i\kappa)}\,\delta_{ll'} \delta_{\vec{m}\vec{m}'} \int\limits_0^\infty dr\ r^{i\kappa+i\omega} \left(\frac k2 \right)^{i\kappa} \sqrt{k} \, J_{\frac{d-3}{2}+l}(kr)\nonumber\\
    & = \frac{e^{\frac{\pi\kappa}{2}}}{2\pi\sqrt{ k\kappa}}\,\Gamma(1-i\kappa) \left(\frac{k}{2}\right)^{-i\omega} \, \frac{\Gamma\left(\frac{\frac{d-1}{2}+l+i(\kappa+\omega)}{2}\right)}{\Gamma\left(\frac{\frac{d-1}{2}+l-i(\kappa+\omega)}{2}\right)} \ \delta_{ll'}\,\delta_{\vec{m}\vec{m}'}\ ,
\\[8pt]
    \beta_{\omega l\vec{m},kl'\vec{m}'} & = - e^{-\pi\kappa}\, \alpha_{\omega l\vec{m},kl'\vec{m}'} \ .
\end{align}

Let us  conclude our discussion of the commutative de~Sitter space with a remark about the two-dimensional space. In $d=2$,  modes \eqref{solutions-(eta,y)-coords} are usually rewritten in terms of Legendre functions, which is the form  we shall use below:
\begin{equation}\label{2-y-modes}
      v_{\omega,\kappa}(\upeta,\rho) 
      = c_{\omega,\kappa} \, (-\upeta)^{-i\omega} (\rho^2 - 1)^{-\frac{i\omega}{2}} Q^{-i\omega}_{-\frac12+i\kappa}(\rho)\, ,\qquad \rho>0\ .
\end{equation}
The variable $\rho = \ty\,$ can
in two dimensions acquire  negative values, and we need to extend \eqref{2-y-modes} to  $\rho<0$. The simplest choice is to extend $\,Q^{-i\omega}_{-\frac12+i\kappa}$ as an even function. The normalisation constant then reads
\begin{equation}\label{normalisation-constant}
    c_{\omega,\kappa}=   \frac{\Gamma(1+i\kappa)}{2\pi \sqrt \kappa}\, \,\frac {e^{-\pi\omega}}{\,\Gamma\left(\frac12+i\kappa-i\omega\right)}\,, \quad |c_{\omega,\kappa}|^2 = \frac{1}{4\pi^2 \sinh(\pi\kappa)}\, e^{-2\pi\omega} \cosh(\pi(\kappa-\omega))  \ .
\end{equation}
One may consider other extensions of  $\,Q^{-i\omega}_{-\frac12+i\kappa}$, such as the one that uses the formula given in \cite{AS}
\begin{equation}\label{Q}
    Q^{-i\omega}_{-\frac 12 +i\kappa}(-\rho) =\pm i e^{\mp \kappa\pi}\, Q^{-i\omega}_{-\frac 12 +i\kappa}(\rho)\,,
\end{equation}
where the two signs correspond to analytic continuations in the upper or lower half-plane, respectively. Using the extension in the upper half-plane, for the value of the normalisation constant $ c_{\omega,\kappa}\, $ we obtain
\begin{equation}
    c_{\omega,\kappa}=   \frac{\Gamma(1+i\kappa)}{2\pi}\,
\sqrt{\frac{e^{\pi\kappa}}{\kappa \cosh(\pi\kappa)}}   \, \,\frac {e^{-\pi\omega}}{\,\Gamma\left(\frac12+i\kappa-i\omega\right)}\,, \quad
|c_{\omega,\kappa}|^2 = \frac{e^{\pi\kappa}}{2\pi^2 \sinh(2\pi\kappa)}  e^{-2\pi\omega}\, \cosh(\pi(\kappa-\omega)) \ .    \nonumber
\end{equation}
Note that the dependence on $\omega$, which is relevant for the calculation of the propagator, is in both cases the same. In the following sections we will stick to the simpler choice \eqref{normalisation-constant}.

\section{Fuzzy harmonics: solutions in noncommutative coordinates}
\label{S:Fuzzy harmonics: solutions in noncommutative coordinates}

The purpose of this section is to show how solutions to the Klein-Gordon equation in $(\upeta,\ty^i)$ coordinates, \eqref{solutions-(eta,y)-coords}, may be turned into eigenfunctions of the Laplacian on the fuzzy de~Sitter space. In the first subsection, we shall recall the definition of the fuzzy de~Sitter space from \cite{Buric:2017yes}, its differential geometry and the Laplacian. This is followed by the main result of the section, the construction of eigenfunctions. While parts of the following discussion are formulated for de Sitter space of general dimension, the full construction is given in two cases, fuzzy dS$_4$ and fuzzy dS$_2$.

\subsection{Fuzzy de Sitter spaces and the Laplacian}
\label{ss3.1}

A fuzzy space in the sense of \cite{Madore:2000aq} is defined by two sets of variables, coordinates $\hat x^\mu$ and {momenta} $\hat p_\alpha$, which satisfy  frame relations of the form
\begin{equation}
    [\hat p_\alpha,\hat x^\mu] = e_\alpha^\mu (\hat x)\ .
\end{equation}
The right hand side is expressed solely in terms of $\hat x^\mu$. Roughly, the structure arising from $\{\hat p_\alpha, \hx^\mu\}$ is supposed to be the quantum version of the classical geometry with the vielbein $e_\alpha^\mu(x)$. Coordinates and momenta are either elements of some abstract algebra, or operators acting on a Hilbert space of states. The algebra $\cal{A}$ generated by $\hx^\mu$ is the fuzzy space. Momenta and relations between them define  differential geometry of the fuzzy space, including the Laplacian. For fuzzy de~Sitter spaces, both $\hat p_\alpha$ and $\hat x^\mu$ are constructed from elements of the Lie algebra $\mathfrak{so}(1,d)$, acting in a particular irreducible representation. Before proceeding to give more details, we describe the two examples of interest. 

Several models of noncommutative dS$_2$ and AdS$_2$ spaces have been defined using the algebra $\mathfrak{so}(1,2)$ in the literature, \cite{Madore:1999fi,Jurman:2013ota,Pinzul:2021cjz}. We define the \textit{fuzzy dS$_2\,$} following the prescription of \cite{Madore:2000aq,cho} as it conforms with our general construction of the fuzzy dS$_d$\,: originally, it was introduced to describe the covariant differential calculus on the $h$-deformed hyperbolic plane, \cite{Madore:1999fi}. Denoting the deformation parameter by $\kbar$,\footnote{Noncommutativity of spacetime is measured by  $\kbar$. Existence of another length scale $\ell$ (or $\Lambda$) related to the spacetime curvature allows to discuss different scaling limits, the noncommutative and the flat space limit. As we do not discuss the latter here, we set $\ell=1$ in most of the text.} coordinates $\heta$, $\hx$ of fuzzy dS$_2$ satisfy
\begin{equation}\label{coord-commutator-fds2}
    [\heta,\hx]=i\kbar\heta\ .
\end{equation}
In terms of $\mathfrak{so}(1,2)$, they can be identified with generators of translations and dilations, respectively
\begin{equation}\label{coords-and-momenta-dS2}
     \heta = - i \kbar P, \quad \hx = - i \kbar D\, \ .
\end{equation}
Momenta are given by
\begin{equation}
    \hp_0= D\, , \qquad \hp_1= -P\ ,
\end{equation}
and the non-vanishing frame brackets are 
\begin{equation}\label{frame-relations-2d}
    [\hp_0,\heta] = \heta\,, \quad [\hp_1,\hx] = \heta\ .
\end{equation}
Therefore, the frame elements $e^\mu_\alpha\,$ give the correct metric of the two-dimensional de~Sitter space in Poincar\'e coordinates. Whereas the differential geometry is insensitive to the coordinate commutator \eqref{coord-commutator-fds2}, it very much depends on the commutators between momenta, in the case at hand
\begin{equation}
    [\hp_0,\hp_1] = \hp_1\ .
\end{equation}
Frame relations and momentum commutators define the differential and give derivatives of zero- and one-forms, respectively:
\begin{equation}
    d\heta = \heta \theta^0\,, \quad d\hx =\heta\theta^1\,, \quad\quad d\theta^0 = 0\,, \quad d\theta^1 = -\theta^0\theta^1\ .
\end{equation}
The  co-frame one-forms $\theta^\alpha$ are dual to derivations $e_\alpha = \text{ad}_{p_\alpha}$. They behave much as their commutative analogues -- they commute with coordinates and anticommute between themselves. Proceeding with the algebra of $p$-forms according to the rules of \cite{Madore:2000aq}, one arrives at the action of Laplacian on scalar fields (elements of $\mathcal{A}$),
\begin{equation}\label{Laplacian-fuzzy-dS2}
    \Delta \Phi = -[\hp_0,[\hp_0,\Phi]] +  [\hp_0,\Phi] + [\hp_1,[\hp_1,\Phi]]\ .
\end{equation}

The \textit{fuzzy dS$_4\,$} was defined in \cite{Buric:2017yes} based on the similarity between the realisation of classical de~Sitter space as a hyperboloid in five-dimensional Minkowski embedding space,
\begin{equation}\label{embedding-relation-commutative}
    X^\alpha X_\alpha = \ell^2
    \,,
    \qquad \alpha = 0,1,\dots,4\,,
\end{equation}
and the expression for the quartic Casimir element of $\mathfrak{g}=\mathfrak{so}(1,4)$,
\begin{equation}\label{quartic-Casimir-expression}
    C_4 = W^\alpha W_\alpha\ .
\end{equation}
The $W_\alpha$ are components of the Pauli-Lubanski vector
\begin{equation}\label{W}
     W^\alpha =\dfrac 18 \,\epsilon^{\alpha\beta\gamma\delta\eta}M_{\beta\gamma} M_{\delta\eta}\,, 
\end{equation}
where we use the convention $\epsilon^{01234}=1$. In any irreducible representation of $\mathfrak{so}(1,4)$, the quartic Casimir acts as a constant. Thus, upon identifying $W_\alpha$ and $X_\alpha$, one recovers the embedding equation \eqref{embedding-relation-commutative}, with the appropriate relation between the cosmological constant and the value of $C_4$. A similar observation can be made for any de~Sitter space of even dimension. In $d=3$, the appropriate choice of coordinates in the negatively curved case was suggested in \cite{Buric:2022ton}. For fuzzy dS$_4$, following the above analogy, the coordinates are chosen among components of the Pauli-Lubanski vector,\footnote{While the definition of coordinates is motivated by the embedding relation, it is the noncommutative Poincar\'e coordinates \eqref{NC-coordinates-4d}, not the embedding ones, that define our fuzzy geometry.}
\begin{equation}\label{NC-coordinates-4d}
    \hat\eta = -\ell (W_0 - W_4), \quad \hx^i = \ell W^i, \qquad i=1,2,3\ .
\end{equation}
Momenta $\hat p_\alpha$ are defined as certain generators of $SO(1,4)$, similarly as in the two-dimensional case. The momenta read 
\begin{equation}
    \hp_0 = D, \quad \hp_i = - P_i\, , \ \qquad i=1,2,3\ .
\end{equation}
Correctness of this choice is verified by the frame relations
\begin{equation}\label{frame-relations}
    [\hp_\mu,\hx^\nu] = \delta_\mu^\nu \,\hat\eta   \, , \ \qquad \mu,\nu=0,1,2,3\,,
\end{equation}
which correspond to the vielbein of the dS$_4$ metric. The relations \eqref{frame-relations} mean that $\hx^\mu = (\heta,\hx^i)\,$ should be thought of as quantisations of Poincar\'e coordinates $(\upeta,\tx^i)$. In all these expressions, elements of the universal enveloping algebra $U(\mathfrak{g})$, are regarded as operators acting in some irreducible representation of $\mathfrak{g}$. The choice of this representation is a part of the quantisation procedure and will be detailed below. However, for purposes of the present discussion, we do not need detailed knowledge of the representation.

As in two dimensions, the frame formalism provides a systematic way to define differential geometry based on \eqref{frame-relations} and the algebra satisfied by the momenta. In the case at hand, this is the Lie algebra
\begin{equation}\label{momentum-momentum-brackets}
    [\hp_0,\hp_i] = \hp_i, \qquad [\hp_i,\hp_j] = 0\ .
\end{equation}
Consequently, we say that fuzzy de~Sitter spaces are of Lie-algebra type. Both two- and four-dimensional spaces share some important properties with their commutative counterparts -- they have constant curvature and satisfy Einstein's equations with a positive cosmological constant, \cite{Brkic:2021lre}. The standard procedure leads to the construction of the Riemannian Laplace-Beltrami operator, concretely
\begin{equation}\label{Laplacian-fuzzy-dS}
    \Delta \Phi = -[\hp_0,[\hp_0,\Phi]] + 3 [\hp_0,\Phi] + [\hp_i,[\hp_i,\Phi]]\ .
\end{equation}
The Laplacian is an operator acting on the algebra generated by noncommutative coordinates. A related operator acting on the Hilbert space of the underlying representation of $SO(1,4)$, termed the ‘quantum-mechanical Laplacian', was analysed in \cite{Brkic:2021lre}. It is possible to write eigenfunctions of \eqref{Laplacian-fuzzy-dS} in terms of eigenfunctions of the quantum-mechanical Laplacian and representation-theoretic quantities such as $6j$-symbols of $SO(1,4)$. We shall follow a much simpler and more direct approach, which makes use of a special set of coordinates on fuzzy dS$_d$: this is the subject of the next subsection. 

{\bf Remark\,} The commutative analogue of the frame introduced above is the set of vector fields
\begin{equation}\label{commutative-frame}
    e_0 = \upeta \partial_{\upeta}, \quad e_i = \upeta \partial_{\tx^i}\ .
\end{equation}
As we have seen, on the noncommutative space, the frame derivations are generated by elements of $\mathfrak{so}(1,d)$. In the commutative setup, this is not the case -- the classical frame \eqref{commutative-frame} does not consist of Killing vectors. However, there is a diffeomorphism that affords a transition between vectors $\{e_0,e_i\}$ and $\{P_0,P_i\}$, where the latter are generators of $SO(1,d)$ that acts on de Sitter space by isometries, \eqref{isometries-dSd-1}-\eqref{isometries-dSd-2}. Indeed, if $\Phi$ is the involution
\begin{equation}\label{diffeomorphism-Phi}
    \Phi: \text{dS}_d \to \text{dS}_d, \qquad (\upeta,\tx^i) \mapsto \left(\upeta^{-1},\frac{\tx^i}{\upeta}\right)\,,
\end{equation}
then the push-forward by $\Phi$ carries $\{e_0,e_i\}$ to $\{P_0,P_i\}$,
\begin{equation}
    \Phi^\ast e_0 = P_0, \quad \Phi^\ast e_i = P_i\ . 
\end{equation}
In the case of the $h$-deformed hyperbolic plane, similar structure was pointed out in \cite{Madore:2000aq}. It implies in particular that, while the algebra of noncommutative coordinates carries a representation of the de Sitter group $SO(1,d)$, its action on the coordinates gets ‘twisted' by $\Phi$. This twisting, however, does not prevent the fuzzy space from having the correct commutative limit.

\subsection{Solutions to the Klein-Gordon equation}

In order to discuss dynamics of fields on the fuzzy de~Sitter space, we first need to solve the equation of motion for a free field $\Phi(\hx^\mu)$. In $d=4$ it is
\begin{equation}\label{Klein-Gordon}
 -[\hp_0,[\hp_0,\Phi]] + 3 [\hp_0,\Phi] + [\hp_i,[\hp_i,\Phi]]+M^2\Phi =0\,,
\end{equation}
$M$ is the mass of the field. The general solution to an operator equation like \eqref{Klein-Gordon} is, as a rule, difficult to find, even in spacetimes with high degree of symmetry built upon a Lie algebra. An exception are equations that reduce to Casimir conditions, e.g. like the eigenvalue equation for the Laplacian on the fuzzy sphere, \cite{Madore:1991bw}. On the four-dimensional fuzzy de~Sitter space we are most interested in, a part of the problem is that commutators between coordinates $\heta$, $\hx^i$ are quite complicated \cite{Buric:2017yes}, so it is hard to compute the left hand side of \eqref{Klein-Gordon} except for the simplest functions $\Phi(\hx^\mu)$. An approach that we shall follow here is to work with a specific choice and ordering of coordinates (similar to the normal ordering in quantum field theory) and try to solve the equation within the set of ordered functions. The result that we obtain is quite interesting: the set of solutions is ‘the same' as the set of modes \eqref{solutions-(eta,y)-coords} found on the commutative de~Sitter space. 

To explain the idea, let us first consider the Klein-Gordon equation on fuzzy dS$_2$. The solutions (up to a Wick rotation) in $(\heta,\hx)$ coordinates were discussed in \cite{Madore:1999fi}; the authors were able to solve the Laplace equation by making use of the special form of the algebra which allows reordering of coordinates for particular functions, for example $ \, \, \heta \, f(\hx) = f(\hx+i\kbar)\, \heta\, $, or $\, \,g(\heta)\,  \exp({ik\hx} ) = \exp({ik\hx})\,g(e^{-\kbar k}\heta)\,$. We follow a different path. Introduce the coordinate $\hy$ by
\begin{equation}
    \hy = \frac 12 \,(\heta^{-1} \hx + \hx\,\heta^{-1})\, ,\qquad [\heta,\hy]=i\kbar  \ .
\end{equation}
This is analogous to the coordinate $\ty$ we considered in the commutative setup. The symmetrisation ensures that $\hat y$ is hermitian. We find 
\begin{equation}\label{py-commutators-2}
\begin{array}{ll}
   [\hp_0,\heta]=  \heta\,,  \quad  &  [\hp_1,\heta]= 0\,, \\[8pt]
   [\hp_0,\hy]= - \hy\,,   & [\hp_1,\hy]= 1  \ .
\end{array}
\end{equation}
These commutation relations imply that frame derivations do not change the ordering of $\heta$ and $\hy$. Therefore if we fix the ordering of variables  $\heta$ and $\hy$, we can solve the Klein-Gordon equation exactly by separation of variables, as in the commutative case. Let us see this in detail. Assuming that the solution is of the form
\begin{equation}
    \Phi(\heta,\hy) = g(\heta)f(\hy) = (-\heta)^{-i\omega} f(\hy)\ ,
\end{equation}
we find that the Klein-Gordon equation reduces to\footnote{The $\dot g$, $\ddot g$ and $f^\prime$, $f^\pprime$ denote the first and second derivatives of functions $g$, $f$ with respect to their arguments.}
\begin{eqnarray}\label{3}
  &&  g(\heta)(1-\hy^2)\,f^\pprime(\hy)-\heta^2\,\ddot{g}(\heta)\,f(\hy)+2\heta\,\dot{g}(\heta)\,\hy f^\prime(\hy)-2 g(\heta)\, \hy  \,f^\prime(\hy)+M^2 g(\heta) f(\hy) \nonumber \qquad\qquad \\[4pt]
  &&\  = (-\heta)^{-i\omega}\Big( (1-\hy^2) f^\pprime(\hy) -(2+2i\omega)\hy f^\prime(\hy) -\big(i\omega(i\omega+1)-M^2  \big) f(\hy) \Big) =0 \,,
\end{eqnarray}
that is, to equation \eqref{KG-spherical}. Therefore, the solution to the equation in the second line of \eqref{3} is $ \, (\hy^2 - 1)^{-i\omega/2}\, Q^{-i\omega}_{-1/2+i\kappa}(\hy)\, $. The modes of the noncommutative scalar field are given by the equivalent of \eqref{2-y-modes},
\begin{equation}\label{2-modes}
      \hat v_{\omega,\kappa}(\heta,\hy ) = c_{\omega,\kappa} \, (-\heta)^{-i\omega} (\hy^2 - 1)^{-\frac{i\omega}{2}} \, Q^{-i\omega}_{-\frac12+i\kappa}(\hy) \ .
\end{equation}

In four dimensions we proceed similarly. We introduce coordinates
\begin{equation}\label{NC-ala-Descartes}
    \hy^i = \frac12 (\hat\eta^{-1} \hx^i + \hx^i \hat\eta^{-1})\ .
\end{equation}
The advantage of $(\heta,\hy^i)$ coordinates again lays in their commutation relations with  momenta,
\begin{equation}\label{py-commutators}
\begin{array}{ll}
   [\hp_0,\heta]=  \heta\,,    &  [\hp_i,\heta]= 0\,,      \\[6pt]
   [\hp_0,\hy^j]= - \hy^j\,,   & [\hp_i,\hy^j]= \delta^j_i\,, \\[6pt]
   [\hp_0,(\hy^j)^n]= -n(\hy^j)^n\,, \qquad  & [\hp_i,(\hy^j)^n] = n (\hy^j)^{n-1}\, \delta^j_i \ .
\end{array}
\end{equation}
These relations imply that frame derivations of a function of one coordinate depend only on that same coordinate, $\, [\hp_\alpha , f(\hy^\mu)] = g(\hy^\mu)$. In turn, this means that, if we impose some particular ordering, the action of the frame derivations, and consequently of the Laplacian, will not change it. Thus we can solve the Klein-Gordon equation by separation of variables. 

Some words of caution are in order. The separation of variables which led to solutions \eqref{solutions-(eta,y)-coords} was done in the spherical polar coordinates $(\upeta,\rho,\theta,\varphi)$. There is no obvious or simple way to quantise the angular coordinates $\theta$ and $\varphi$. Instead, we start by re-expressing the commutative solutions using ‘Cartesian' coordinates $(\upeta,\ty^i)$. This turns out not to be too difficult. Let us, for the sake of the argument, analyse the spatial part of the solution \eqref{KG-sferne} which is finite at $\rho=0$.\footnote{The same analysis can be done for the basis of functions \eqref{solutions-(eta,y)-coords} that we actually use: then instead of $\rho$ we would use $\rho^{-1}$ and expand solution around $\rho^{-1}=0\,$. Alternatively,  we can use one of the contiguous relations between the hypergeometric functions to express \eqref{solutions-(eta,y)-coords} through \eqref{KG-sferne}, then proceed with the Taylor expansion in $\rho$.} It contains two factors: $\rho^l\, Y_l^m(\theta,\varphi)$  and $\, _2F_1( a,b ;\,  3/2+l  ;\rho^2)$. The first factor is a homogeneous polynomial of $\ty^i$ of degree $l$, while the second can be expanded in Taylor series in $\rho^2=\ty^i\ty_i$. Thus, the spatial part of the solution expands as a series in $\ty^i$. 
   
Let us impose the operator ordering as
\begin{equation}
    \Phi(\heta,\hy^i) = \phi(\heta)\, f(\hy^1)\, g(\hy^2)\, h(\hy^3)\,,
\end{equation}
and denote the ordered monomials of spatial coordinates  by $\hat f_{n_1,n_2,n_3}$,
\begin{equation}\label{monom}
    \hat f_{n_1,n_2,n_3}=(\hat y^1)^{n_1} (\hat y^2)^{n_2} (\hat y^3)^{n_3}\ . 
\end{equation}
From (\ref{Laplacian-eta-y}) we find that the commutative Laplacian $\Delta_{dS_4}$, acting on functions of the form
\begin{equation}\label{ansatz}
    v(\upeta,\ty^i) =  (-\upeta)^{-i\omega} F(\ty^i)\,,
\end{equation}
reduces to operator
\begin{equation}\label{KG-Cartesian}
    \Delta^{-i\omega}_{dS_4} =    \p_{\ty^i}\p_{\ty_i} -\ty^i\ty^j\p_{y^i} \p_{y^i} \p_{y^j}  -(2i\omega +4)\,\ty^i\p_{\ty^i} -3i\omega +\omega^2  \ .
\end{equation}
We have, further,
\begin{align}
& \ty^i\p_{\ty^i}\,f_{n_1,n_2,n_3} =(n_1+n_2+n_3) \, f_{n_1,n_2,n_3} \,, \\[4pt]
&  \p_{\ty^i}\p_{\ty_i}\, f_{n_1,n_2,n_3}= n_1(n_1-1) f_{n_1-2,n_2,n_3} +n_2(n_2-1)f_{n_1,n_2-2,n_3} + n_3(n_3-1) f_{n_1,n_2,n_3-2}  \ . \\
 \intertext{In parallel, from  commutation relations \eqref{py-commutators} we compute}
  & [\hp_0,\hat f_{n_1,n_2,n_3} ] = -\, (n_1+n_2+n_3)\, \hat f_{n_1,n_2,n_3} \,,\\[2pt]
& [\hp^i,[\hp_i,\hat f_{n_1,n_2,n_3}] ] = n_1(n_1-1) \hat f_{n_1-2,n_2,n_3} + n_2(n_2-1)\hat f_{n_1,n_2-2,n_3} + n_3(n_3-1) \hat f_{n_1,n_2,n_3-2}\ .
\end{align}
Clearly, on functions of spatial coordinates, commutators $ [\hp_i,\ ] $ act as $\p_{\ty^i}\,$, and  commutator $\, [\hp_0,\ ] $ as $ \,- \ty^i\p_{\ty^i}\,$ - this follows by applying the commutators to Taylor-expanded functions. Furthermore, one verifies that the same substitutions map the fuzzy Laplacian to the commutative one. This implies that if we write the solutions in the form 
\begin{equation}\label{quantisation-of-solutions}
    \hat F(\hy^i)  =\sum \hat c_{n_1,n_2,n_3}\, \hat f_{n_1,n_2,n_3} \,,\qquad F(\ty^i)  =\sum c_{n_1,n_2,n_3}\,  f_{n_1,n_2,n_3} \ ,
\end{equation}
the equations for coefficients $\, \hat c_{n_1,n_2,n_3}$ and $\,  c_{n_1,n_2,n_3}$ are identical, and read
\begin{eqnarray} \label{KG-coeff}
  &&  (n_1+2)(n_1+1)\, \hat c_{n_1+2,n_2,n_3} +(n_2+2)(n_2+1)\, \hat c_{n_1,n_2+2,n_3} + (n_3+2)(n_3+1) \, \hat c_{n_1,n_2,n_3+2} \qquad \nonumber \\[6pt]
  &&\qquad = \Big( (n_1+n_2+n_3)^2 +(2i\omega+3)(n_1+n_2+n_3) +(-\omega^2+3i\omega +M^2)\Big) \, \hat c_{n_1,n_2,n_3} \ .             
\end{eqnarray}
Although the last equation looks complicated, we have already found its solution: $\, \hat c_{n_1,n_2,n_3} \,$ are given by the Taylor expansion of \eqref{KG-sferne}. 

We end this section with some remarks. One readily notices that the relation for the coefficients \eqref{KG-coeff} would still be obtained if we used some other ordering of coordinates, e.g. 0321 instead of 0123. Any ordering gives a set of solutions to the Klein-Gordon equation, and due to complicated commutation relations among $\hy^i$-s, it is unclear how these different sets of solutions are related to each other. The solutions however should all have the same commutative limit: this follows from the fact that the commutators of coordinates vanish in this limit.\footnote{This qualitative argument will be more precisely discussed in terms of operator kernels in the next section.} This result ensures that on the macroscopic scale (i.e. in the commutative limit), fields on fuzzy de~Sitter space behave classically.

Another observation related to  the given procedure is that we do not know whether it gives all i.e. complete set of eigenfunctions of the noncommutative Laplacian. To answer this question, we need to rewrite the Klein-Gordon equation as a matrix or differential equation. 

Both of the raised points may be addressed by substituting for coordinates and momenta concrete operators acting in a unitary irreducible representation of $SO(1,d)$. In this way, the fuzzy harmonics are turned into matrices, not unlike fuzzy spherical harmonics on the fuzzy sphere. Due to non-compactness of $SO(1,d)$, these matrices are infinite-dimensional and we prefer to write them as integral kernels. This is the subject of the next section.

\section{All fuzzy harmonics: integral kernels}
\label{S:All fuzzy harmonics}

In the present section we shall obtain a complete set of modes of the fuzzy scalar field. To this end, in the first subsection we will show how to write functions on the fuzzy de Sitter space as integral kernels depending on $\,2(d-1)$ commuting real variables. The kernels that we use  are matrix elements  written in an eigenbasis of the conformal time coordinate. Solutions of the last section are then analysed in this basis. In the second subsection, we use the integral-kernels formulation to find the most general solution to the fuzzy Klein-Gordon equation. In four dimensions, there are infinitely many more modes than in the commutative case and they are organised into a discrete Kaluza-Klein tower.

\subsection{Harmonics in the eigenbasis of conformal time}

Recall that coordinates and momenta of  fuzzy dS$_d$ are defined as operators acting in a certain representation of $SO(1,d)$. Up to this point, we did not need much information about this representation. However, the coordinate spectra, number of degrees of freedom etc. are all expected to depend non-trivially on it. For both fuzzy dS$_2$ and fuzzy dS$_4$, our construction is based on the principal series representations of $\mathfrak{so}(1,d)$.\footnote{The principal series representation is singled out by the condition of positivity of the cosmological constant, \cite{Corfu}.} The latter may be realised on spaces of conformal fields in $\,(d-1)\,$ dimensions. More precisely, we can write the generators of $SO(1,d)$ as differential operators acting on functions $\psi_a(z_i)$ of $\,(d-1)\,$ real variables $z_i$,
\begin{align}\label{conformal-generators-1}
    & P_i = - \partial_i, \quad L_{ij} = z_i \partial_j - z_j \partial_i - \Sigma_{ij}\,,\\[2pt]
    & D = -z^i \partial_i - \mathbb{\Delta}, \quad K_i = - z^2 \partial_i - 2z_i D - 2 z^j \Sigma_{ij}\ .\label{conformal-generators-2}
\end{align}
The functions $\psi_a(z_i)$ take values in some finite-dimensional representation space $V$ of $SO(d-1)$ and $\Sigma_{ij}$ are matrix representatives of $L_{ij}$ in this representation. Furthermore, $\,\mathbb{\Delta}\,$ is a constant and unitarity requires $\mathbb{\Delta} = (d-1)/2+i\tau$ with $\tau\in\mathbb{R}$. Indices $\,a\,$ run over a basis $\{e^a\}$ for $V$ and are raised and lowered by an $SO(d-1)$-invariant metric on it. The Hilbert space of functions $\mathbb{R}^{d-1} \to V $ will be denoted by $\mathcal{H}$. The inner product on $\mathcal{H}$ is given by the integration with respect to the standard measure on $\mathbb{R}^{d-1}$,
\begin{equation}
    (\psi_1,\psi_2) = \int_{\mathbb{R}^{d-1}} d^{d-1} z\ \left(\psi_1^a(z^i)\right)^\ast \psi_{2a}(z^i)\ .  
\end{equation}
If we denote the principal series representation by $\pi_{\mathbb{\Delta},V}$, the representations $\pi_{\mathbb{\Delta},V}$ and $\pi_{d-1-\mathbb{\Delta},V}$ are isomorphic to one another.\footnote{This statement is true if $V$ is a symmetric traceless tensor representation, the only case we shall consider. More generally, $\pi_{\mathbb{\Delta},V}\cong\pi_{d-1-\mathbb{\Delta},V^R}$, where $V^R$ is the reflected representation of $V$, see \cite{Kravchuk:2018htv} for details.} For fuzzy dS$_2$, there is no rotation group $SO(d-1)$, and matrices $\Sigma_{ij}$ are set to zero. In the $d=4$ case, the fuzzy dS$_4$ is defined using the spin-$1/2$ principal series representation, for which\footnote{The spinor representation $s=1/2$ is used because for $s=0$ the Casimir $C_4$ vanishes, which corresponds to the infinite cosmological constant, \cite{Corfu}. In fact, $C_4=0$ implies $W^\alpha=0$.}
\begin{equation}\label{4d-representation}
    \Sigma_{ij} = -\frac{i}{2}\,\epsilon_{ijk}\sigma_k\ .
\end{equation}
For more details on unitary representations of $SO(1,d)$, we refer the reader to \cite{Dobrev:1977qv}, in $d=2\,$ to \cite{Bargmann:1946me}.

States of the fuzzy geometry are represented by wavefunctions $\psi_a(z_i)$ and coordinates by differential/integral operators acting on them. E.g. for the fuzzy dS$_2$, the coordinates are
\begin{equation}
     \heta = i \kbar \partial_z\,, \quad \hx =  i \kbar (z\partial_z + \mathbb{\Delta})\ .
\end{equation}
The formulation in terms of wavefunctions allows to study spectra of coordinates, overlaps of their eigenstates, expectation values etc. using standard methods of quantum mechanics. The mentioned observables reveal physical properties of our model. In the remainder of this subsection, we shall focus on the matrix elements of fuzzy harmonics in the eigenbasis of the time coordinate.

Since our definition of the coordinate $\hat y\,$ involves the inverse of the derivative operator $\heta$, it is convenient to pass to Fourier space, where $\heta$ and $\heta^{-1}$ are represented by multiplication operators,
\begin{equation}
    P = - i q, \quad D = q \partial_q + 1 - \mathbb{\Delta}, \quad K = i q \partial_q^2 - 2i D \partial_q\ .
\end{equation}
In other words, we work in the basis of eigenstates of $\heta$. In this basis, the other coordinate $\hat\rho=\hy\,$ is represented by a derivative-like operator, 
\begin{equation}\label{radis-op-fdS2}
    \heta = -\kbar q, \qquad \hat\rho = \hy = \frac12 \left(\heta^{-1} \hx + \hx \heta^{-1}\right) = i \partial_q - i \, \frac{\,\mathbb{\Delta}-\frac12\, }{q}  \ .
\end{equation}
Coordinate $\hat\rho$ has continuous spectrum $(-\infty,\infty)$. Its eigenstates, $\,\hat\rho|\lambda\rangle = \lambda |\lambda\rangle$, found by solving a differential equation, read
\begin{equation}\label{lambda-momentum-space}
    \langle q|\lambda\rangle = \frac{1}{\sqrt{2\pi}}\, q^{-\frac 12+\mathbb{\Delta}}\, e^{-i \lambda q} \ .
\end{equation}
They satisfy orthogonality relations
\begin{equation}
      \int\limits_{-\infty}^\infty dq \, \langle \lambda|q\rangle\langle q|\lambda'\rangle =\delta(\lambda-\lambda')\, ,\qquad
    \int\limits_{-\infty}^\infty d\lambda \, \langle q|\lambda\rangle\langle \lambda|q'\rangle =\delta(q-q') \ .
\end{equation}
The eigenstates $|\lambda\rangle\, $ are all we need in order to find matrix elements of fuzzy harmonics. Indeed, let $\,v_{\omega,\kappa}(\upeta,\ty)$ be a solution to the Klein-Gordon equation on the commutative dS$_2$ obtained in Section \ref{S:Coordinates on dSd field quantisation and vacua} and $\,\hat v_{\omega,\kappa}(\heta,\hy)$ the corresponding fuzzy solution found in the last section. We find
\begin{align}
\hat  v_{\omega,\kappa}(\eta_L,\eta_R) & = \langle \eta_L|\hat v_{\omega,\kappa}|-\eta_R\rangle = c_{\omega,\kappa} \int d\lambda\ \langle \eta_L | (-\heta)^{-i\omega}(\hy^2-1)^{-\frac{i\omega}{2}} Q^{-i\omega}_{-\frac12+i\kappa}(\hy) |\lambda\rangle \langle\lambda| -\eta_R \rangle\nonumber\\[4pt]
    & = \frac{c_{\omega,\kappa}}{\sqrt{2\pi\kbar}}\left(\frac{\eta_R}{\kbar}\right)^{-i\tau} \int d\lambda\ (\lambda^2-1)^{-\frac{i\omega}{2}} Q^{-i\omega}_{-\frac12+i\kappa}(\lambda) \langle \eta_L | (-\heta)^{-i\omega}  |\lambda\rangle e^{\frac{i\lambda\eta_R}{\kbar}} \label{v-matrix}  \\[4pt]
    & = \frac{c_{\omega,\kappa}}{2\pi\kbar} \left(\frac{\eta_R}{\kbar}\right)^{-i\tau} \, (-\eta_L)^{-i\omega} \left(\frac{-\eta_L}{\kbar}\right)^{i\tau} \int d\lambda\ e^{\frac{i\lambda(\eta_L+\eta_R)}{\kbar}} (\lambda^2-1)^{-\frac{i\omega}{2}} Q^{-i\omega}_{-\frac12+i\kappa}(\lambda)\label{v-matrix-integral} \\[4pt]
     & = C\, \frac{c_{\omega,\kappa}}{2\pi \kbar^{i\omega+1/2}} \,(-\eta_L)^{i\tau-i\omega} \eta_R^{-i\tau} \, (\eta_L+\eta_R)^{i\omega-\frac12}\, J_{i\kappa}\left(\frac{\eta_L + \eta_R}{\kbar}\right)\ .\label{v-matrix-element}
\end{align}
The minus sign in the state $|-\eta_R\rangle$ is due to our choice of conventions summarised in Appendix \ref{A:Representation-theoretic background}. By inserting the identity $\,\mathbb{1} = |\lambda\rangle\langle\lambda|\,$ and using the expression for $|\lambda\rangle$ in Fourier space, we get an integral representation for the matrix element. The final result is obtained by a regularised Fourier transform, as outlined in Appendix \ref{A:Fourier transformations}, where we also give the value of the constant $C$. One can readily check that the final result \eqref{v-matrix-element} satisfies the Klein-Gordon equation in Fourier space. The field modes \eqref{v-matrix-element} allow to write explicitly the quantisation map and its inverse, $f(\upeta,\ty)\longleftrightarrow \hat f(\eta_L,\eta_R)\,$.

In four dimensions one can follow much the same path, although the computations become more involved. We work with a Fourier space realisation of the principal series, similarly as above, with
\begin{align}\label{conformal-generators-Fourier-1}
    & P_i = - i q_i \, , \quad L_{ij} = q_i \partial_{q^j} - q_j \partial_{q^i} - \Sigma_{ij}\,,\\[4pt]
    & D = q^i \partial_{q^i} + 3 - \mathbb{\Delta}, \quad K_i = iq_i \partial^2 - 2i D \partial_{q^i} - 2i\Sigma_{ij}\partial_{q_j}\ .\label{conformal-generators-Fourier-2}
\end{align}
We have denoted the Fourier variables dual to $z_i$ by $q^i$. In this realisation, the Poincar\'e coordinates on the fuzzy dS$_4$ are represented by the operators
\begin{equation}\label{heta}
    \heta = -\frac12 \, \sigma^i q_i = -\frac q2  \,\Pi \, , 
    \qquad \hx_j = -\frac{i}{2}\, \sigma^k q_k \partial_{q^j} + i\, \frac{\,\mathbb{\Delta}-2\,}{2}\,\sigma_j\ .
\end{equation}
We have introduced $\,\Pi ={\sigma^i q_i}/{q}\, $, a notation that shall be useful throughout the section. The Fourier space is again closely related to the eigenbasis of the time coordinate. Since $\heta$ is a scalar operator, its eigenstates may be chosen with well-defined rotational quantum numbers. Such eigenstates are
\begin{equation}\label{eigenstates-of-eta}
  \langle q,\vartheta,\phi  |\,\eta,j,m_j, \mathrm{a}\rangle =  \dfrac{1}{\sqrt 2\, \eta} \, \delta\Big(q- (-1)^\mathrm{a} 2 \eta\Big) \,\varphi^\mathrm{a}_{j,m_j}(\vartheta,\phi) \,, \quad\quad (-1)^\mathrm{a} \equiv\left\{\begin{array}{ll}
     -1, &  \mathrm{a}=\uparrow     \\[2pt]
       1 , & \mathrm{a}=\downarrow 
    \end{array}   \right. \,,
\end{equation}
where $(q,\vartheta,\phi)$ are the spherical polar coordinates constructed out of $ q^i$, $q^2 = q_i q^i$. The $\,\varphi^{\mathrm{a}}_{j,m_j}$ are spinor harmonics: $j$ is a half-integer, $j=\frac 12,\frac 32,\dots \,$ and $\, m_j=-j,-j+1,\dots,j\,$. Given a fixed pair $(j,m_j)$, there are two linearly independent harmonics, which we index by $\mathrm{a}$.\footnote{For any fixed $\mathrm{a}$, the $\,\varphi^{\mathrm{a}}_{j,m_j}$ is a two-component object, whose index labelling these components has been suppressed.} With the choice of basis from \cite{Bjorken:100769}, $\mathrm{a}$ runs over $\{+,-\}$; the explicit expressions for spinor harmonics are collected in Appendix \ref{A:Special functions}. We shall mostly work with the basis that diagonalises the operator $\Pi$,
\begin{equation}
\varphi^{\uparrow,\downarrow}_{j,m_j}=\frac{1}{\sqrt 2}(\varphi^{+}_{j,m_j}\pm \varphi^{-}_{j,m_j})\,, \ \qquad \Pi \varphi^{\uparrow}_{j,m_j}=  \varphi^{\uparrow}_{j,m_j} ,\quad \Pi \varphi^{\downarrow}_{j,m_j}= - \varphi^{\downarrow}_{j,m_j}\ .
\end{equation}
Again, it is possible to find eigenstates of the radial coordinate $\hat\rho^2$ in the eigenbasis of $\hat\eta$. To this end, we start with the expressions for $\hy^i$ and $\hat\rho^2$ in Fourier space,
\begin{eqnarray}
 &&   \hy_i = i\p_{q^i} - i\Big(\mathbb{\Delta}-\frac32\Big) \frac{q_i}{q^2} + \frac 12\, \epsilon_{ijk}\, \frac{q^j}{q^2}\, \sigma^k  \, , \\[4pt]
 && \hat \rho^2 = -\p_{q_j}\p_{q^j} +\frac{2\mathbb{\Delta}-3}{q^2}\, q^i\partial_{q^i} 
 +\frac{2i}{q^2}\,\epsilon^{ijk} q_j \partial_{q^k} \sigma^i -\Big( (\mathbb{\Delta}-2)^2-\frac 34 \Big)\frac{1}{q^2}\ .  \nonumber 
\end{eqnarray}
The eigenstates of  $\,\hat\rho$, normalised to $\delta(\lambda-\lambda')$, are obtained by solving a differential equation, and read
\begin{equation}
     \langle q,\vartheta,\phi\, |\lambda,j,m_j,{\mathrm{a}}\rangle = {\sqrt\lambda}\, q^{\mathbb{\Delta}-2}\, J_{\sqrt{j(j+1)}}\,(\lambda q)\, \varphi^{\mathrm{a}}_{j,m_j}(\vartheta,\phi)\ , \qquad \lambda>0 \ .
\end{equation}
Since both $\heta\,$ and $\hat\rho\,$ are scalars, their overlaps are diagonal in the rotational quantum numbers $(j,m_j,\mathrm{a})$. We find
\begin{equation}
    \langle \eta, j, m_j, \mathrm{a}
    \,| \lambda, j', m_{j'}, \mathrm{a'}\rangle = \delta_{jj'}\,\delta_{m_j m_{j'}}\, \delta_{\mathrm{aa'}} \ \frac{\sqrt \lambda}{\sqrt 2\, \eta}\,\big( (-1)^\mathrm{a}\, 2\eta \big)^{\mathbb{\Delta}} \,\, J_{\sqrt{j(j+1)}}\big( (-1)^\mathrm{a}\, 2\eta  \lambda\big) \ .
\end{equation}
Following the same procedure as in two dimensions, we can obtain an integral representation for matrix elements of fuzzy harmonics $\,\hat v_{\omega,l,m,\kappa}$. We have seen in the last section that several prescriptions for defining these functions, that make use of different operator orderings, are available. The operator ordering, however, does not effect our definition of scalar modes with $\,l=m=0$. We focus on these for the remainder of the subsection,
\begin{align}
 \hskip-10pt   \hat v_{\omega,0,0,\kappa} &= (-\heta)^{-i\omega}  \, \hat\rho^{-\frac32-i(\kappa+\omega)}\ _2F_1\Big(\frac{3+ 2i(\kappa+\omega)}{4},\frac{1+2i(\kappa+\omega)}{4};1+i\kappa;\hat\rho^{-2}\Big)  = v_{\omega,\kappa}(\heta, \hat\rho)\, \hat\rho^{-1}  \ . 
\end{align}
In the last equation, $v_{\omega,\kappa}$ are the two-dimensional modes we encountered before. We find the matrix elements of the above operators between the eigenstates of $\,\heta\,$,
\begin{align}
    & \hat v_{\omega,0,0,\kappa}(\eta_L,j_L,m_{jL},\mathrm{a}_L; \eta_R,j_R,m_{jR},\mathrm{a}_R) = \langle \eta_L,j_L,m_{jL},\mathrm{a}_L|\hat v_{\omega,0,0,\kappa}\,  |\eta_R,j_R,m_{jR},\mathrm{a}_R\rangle    \nonumber  \\[4pt]
  & = \langle \eta_L,j_L,m_{jL},\mathrm{a}_L|\ c_{\omega,\kappa} (-\heta)^{-i\omega} \hat\rho^{-1} (\hat \rho^2-1)^{-\frac{i \omega}{2}}\, Q^{-i\omega}_{-\frac 12+i\kappa}(\hat\rho) |\eta_R,j_R,m_{jR},\mathrm{a}_R\rangle\ . 
\end{align}
By inserting the identity written in the eigenbasis of $\hat\rho$, the dependence on all variables except for the time eigenvalues $\eta_{L,R}$ trivialises,
\begin{align}\label{v-matrix-element-4d}
   & \hat v_{\omega,0,0,\kappa}(\eta_L,\dots,\mathrm{a}_R)  = c_{\omega,\kappa} \sum_{j,m_j,\mathrm{b}} \int  d\lambda\ \langle \eta_L,j_L,m_{jL},\mathrm{a}_L| (-\heta)^{-i\omega} \hat\rho^{-1} (\hat \rho^2-1)^{-\frac{i \omega}{2}}\, Q^{-i\omega}_{-\frac 12+i\kappa}(\hat\rho) |\lambda,j,m_j,\mathrm{b}\rangle \times \nonumber  \\[4pt]
   & \times \langle \lambda,j,m_j,\mathrm{b}\, |\eta_R,j_R,m_{jR},\mathrm{a}_R\rangle = 4c_{\omega,\kappa} \,\delta_{j_L j_R}\,  \delta_{m_{jL} m_{jR}} \,\delta_{\mathrm{a}_L \mathrm{a}_R}((-1)^{\mathrm{a}_L})^{\mathbb{\Delta}+\mathbb{\Delta}^\ast+i\omega}\,\eta_L^{\mathbb{\Delta}-1-i\omega}\, \eta_R^{\mathbb{\Delta}^\ast-1}\times   \\[4pt]
 & \hskip3cm \times \int d\lambda\  (\lambda^2-1)^{-\frac{i \omega}{2}}  \, Q^{-i\omega}_{-\frac 12+i\kappa}(\lambda) \, J_{\sqrt{j_L(j_L+1)}}   \big((-1)^{\mathrm{a}_L}2\eta_L\lambda\big) \, J_{\sqrt{j_R(j_R+1)}}\big((-1)^{\mathrm{a}_R}2\eta_R\lambda\big)\ . \nonumber
\end{align}
The integral in the last line is the analogue of \eqref{v-matrix-integral} in two dimensions. Clearly, the four-dimensional result is more involved. We leave for future work the question of whether the integral, possibly after regularisation, admits a closed form expression.

\subsubsection{Commutative limit}

Let us comment on the manifestation of the commutative limit $\kbar\to 0\,$ in terms of integral kernels. We focus on two dimensions. By a similar calculation as the one done for the modes $\hat v_{\omega,\kappa}$, for any pair of functions $f(\heta)$ and $g(\hy)$ we find the matrix elements of their products
\begin{align}
    \langle q_L| f(\heta) g(\hy) |-q_R\rangle & = \frac{1}{2\pi} \left(\frac{q_L}{-q_R}\right)^{i\tau} \tilde g(q_L + q_R) f(-\kbar q_L)\,,\\
    \langle q_L| g(\hy) f(\heta) |-q_R\rangle & = \frac{1}{2\pi} \left(\frac{q_L}{-q_R}\right)^{i\tau} \tilde g(q_L + q_R) f(\kbar q_R)\ .
\end{align}
Here $\tilde g$ denotes the inverse Fourier transform of $g$. Let us assume that $f$ is an analytic function in a neighbourhood of 0. In that case, by subtracting the last two equations and Taylor-expanding, we get
\begin{equation}
    \langle q_L | [f(\heta),g(\hy)] | -q_R \rangle = -\frac{\kbar}{2\pi} \left(\frac{q_L}{-q_R}\right)^{i\tau} \tilde g(q_L + q_R) (q_L + q_R) f'(0)\  + O\left(\kbar^2\right)\ .
\end{equation}
Therefore, for an appropriate choice of observables $f$ and $g$, that is those for which $f(0)\neq0$, $f$ is analytic at zero, and $g$ admits a Fourier transform, the commutator is correctly suppressed as $\kbar\to 0$. Having discussed the commutative limit, we shall set $\kbar = 1$ for the remainder of the section.

\subsection{General solution to the Klein-Gordon equation}
\label{SS:General solution of the Klein-Gordon equation}

The explicit realisation of the Hilbert space of states of the fuzzy geometry allows us also to discuss the completeness of the set of harmonics analysed above. To this end, consider the fuzzy coordinate algebra $\mathcal{A}$, that of operators on the carrier space of the principal series representation $\pi_{\mathbb{\Delta},V}$. In writing elements $f$ of $\mathcal{A}$, we shall employ the isomorphism
\begin{equation}\label{NC-function-algebra}
    \mathcal{A} = \text{End}(\mathcal{H}) \cong \mathcal{H} \otimes \mathcal{H}^\ast\,,
\end{equation}
and thus regard these elements as vector-valued functions of two sets of variables, $f^a{}_b(z_L^i,z_R^j)$. For principal series representations, $\mathcal{H}^\ast$ is isomorphic to $\mathcal{H}$; we denote $\,\mathbb{\Delta}_R =(d-1)/2+i\tau $, $\ \mathbb{\Delta}_L = \mathbb{\Delta}_R^\ast =  (d-1)/2-i\tau $. In this way of writing, it is manifest that the fuzzy space is $\,2(d-1)$-dimensional, together with some additional discrete structure described by the indices $a,b\,$ that functions $f^a{}_b(z_L^i,z_R^j)$ carry. Our aim for the rest of the section is to quantify this difference in the number of degrees of freedom between commutative and noncommutative cases, which arise whenever $d>2$. 

We shall solve the eigenvalue problem for the Laplacian on the space \eqref{NC-function-algebra}. To this end, recall that elements of the Lie algebra $\mathfrak{so}(1,d)$ act on operators in $\mathcal{A}$ via commutators. When written in terms of kernels $f^a{}_b(z_L^i,z_R^j)$ this action takes the form
\begin{equation}\label{diagonal-action-symmmetry-algebra}
    \text{ad}_X \mapsto X^L + X^R\ .
\end{equation}
In particular, the fuzzy Laplacian \eqref{Laplacian-fuzzy-dS} acts on the kernels via
\begin{equation}\label{NC-Laplacian-general}
    \Delta = -\left(D^L + D^R\right)^2 + (d-1) \left(D^L+D^R\right) + \sum_{i=1}^{d-1} \left(P^L_i + P^R_i\right)^2\ .
\end{equation}
Using expressions for the group generators \eqref{conformal-generators-1}-\eqref{conformal-generators-2}, the Laplacian is turned into a differential operator. Notice that the operator does not depend on the spin of the principal series representation $\mathcal{H}$, as it only involves generators of translations $P_i$ and dilations $D$. In fact, \eqref{NC-Laplacian-general} turns out to be closely related to the Laplacian on the commutative de Sitter space. We proceed to exhibit this relation. Let us introduce variables $y^i$ and $\xi^i$ by
\begin{equation}\label{sums-and-differences}
    z_L^i = y^i + \xi^i, \quad z_R^i = y^i - \xi^i\ .
\end{equation}
The radial coordinate constructed out of $\xi^i$ will be denoted by $\xi^2 = \xi^i \xi_i$. Using \eqref{conformal-generators-1}-\eqref{conformal-generators-2}, we have
\begin{equation}
    P^L_i + P^R_i = -\partial_{y^i}, \quad D^L + D^R = - y^i \partial_{y^i} - \xi^i \partial_{\xi^i} - \mathbb{\Delta}_L - \mathbb{\Delta}_R\,,
\end{equation}
and from here the Laplacian is readily computed. The operator turns out to be invariant under $SO(d-1)$-rotations of the $\xi^i$. Therefore, it can be written as a differential operator in $(\xi,y^i)$ only, namely 
\begin{equation}\label{NC-laplacian-expression}
    \Delta = -y^i y^j \partial_{y^i} \partial_{y^j} + \partial_{y^i} \partial_{y_i} - \xi^2 \partial_\xi^2 - 2\xi\partial_\xi \, y^i \partial_{y^i} - (3d-2)y^i\partial_{y^i} - (3d-2)\xi\partial_\xi\ - 2(d-1)^2.
\end{equation}
 We have used that $\,\mathbb{\Delta}_L + \mathbb{\Delta}_R=d-1$. The operator \eqref{NC-laplacian-expression} commutes with $\xi\partial_\xi$, so it preserves the space of functions of the form
\begin{equation}
  g(\vartheta)(-\xi)^{\delta} f(y^i) \, e_a \otimes e_b \,,
\end{equation}
where $f$, $g$ and $\delta$ are arbitrary. We used $\vartheta$ to denote collectively the angles of the spherical polar coordinates constructed out of $\xi^i$; as a basis for functions on the corresponding $S^{d-2}$ we can take e.g. spherical harmonics. 

We can now make the main observation of the present subsection. From above, one finds the action of the Laplacian on functions of the form $\, (-\xi)^{i\omega-1+d}f(y^i)\,$,
\begin{equation}\label{all-fuzzy}
    \Delta^{i\omega-1+d} = -y^i y^j \partial_{y^i} \partial_{y^j} + \partial_{y^i} \partial_{y_i} - (2i\omega+d)y^i\partial_{y^i} -i\omega (i\omega+d-1)\ .
\end{equation}
Interestingly, comparing this expression to the equation \eqref{Laplacian-eta-y}, we see that, upon identifying $\ty^i$ and $y^i$ (and denoting, by a slight abuse of notation, $\rho^2 = y_i y^i$), \eqref{all-fuzzy} coincides with the action of the commutative Laplacian on dS$_d$. Therefore the solutions to the corresponding equations, after the replacement $\,-i\omega \to i\omega-1+d\,$, are identical, and given by \eqref{solutions-(eta,y)-coords}. The `positive frequency' modes of the scalar field  on fuzzy dS$_d$ are
\begin{align}\label{all}
    w^{l',\vec{m}',ab}_{\omega,l,\vec{m},\kappa}(z_L^i, &z_R^j) =  Y_{l'}^{\vec{m}'}(\vartheta) \, (-\xi)^{i\omega-d+1} \,   \rho^{-\frac{d-1}{2}-i(\omega+\kappa)} \\[4pt]
    & \times\,_2F_1\Big(\frac{2l+d-1 + 2i(\omega+\kappa)}{4},\frac{-2l+5-d+2i(\omega+\kappa)}{4};1+i\kappa;\rho^{-2}\Big)\, Y_l^{\vec{m}}(\theta)  \, e^a \otimes e^b\ .\nonumber
\end{align}
To each commutative solution one associates in this way an $L^2(S^{d-2})\times\text{dim}(V)^2$ worth of noncommutative ones, corresponding to an arbitrary choice of the quantum numbers $l'$, $\vec{m}'$ and the vector $e_a\otimes e_b$. The sphere $S^{d-2}$ is to be thought of as ‘internal space'. Due to its  compactness, the fuzzy harmonics form an infinite discrete tower of Kaluza-Klein modes over the commutative harmonics. Clearly, \eqref{all} gives a complete set of fuzzy harmonics.

\subsubsection{Further details in two dimensions}

Naively, our analysis might suggest to identify ‘almost-classical' solutions $\hat v_{\omega,l,\vec{m},\kappa}$ of Section \ref{S:Fuzzy harmonics: solutions in noncommutative coordinates} with the lowest KK modes $w^{0,0,ab}_{\omega,l,\vec{m},\kappa}$, for some values of indices $a$, $b$. We show here that this is not the case, as may be already seen in two dimensions. Indeed, starting with the modes \eqref{all},
\begin{equation}\label{2-y}
      w_{\omega,\kappa}(\xi,y ) = (-\xi)^{i\omega-1} (y^2 - 1)^{-\frac{i\omega}{2}} \, Q^{-i\omega}_{-\frac12+i\kappa}(y)\,,
\end{equation}
we can perform the Fourier transform to re-express them in terms of variables $\,-\eta_{L,R} = \chi \pm \zeta$,
\begin{align}
    & w_{\omega,\kappa}(\eta_L,\eta_R) =  \iint 2 d\xi\,  dy \, e^{-2i\zeta\xi}\, (-\xi)^{i\omega-1}\, e^{-2i\chi y} \,(y^2-1)^{-\frac{i\omega}{2}}  \, Q^{-i\omega}_{-\frac 12+i\kappa}(y)
       \nonumber  \\
       & = 2  C\,  (-2\chi)^{-\frac 12 +i\omega} J_{i\kappa}(-2\chi) \int\limits_{-\infty}^\infty d\xi\, e^{2i\zeta\xi}\,\xi^{i\omega-1}  = E  \left(\eta_L-\eta_R \right)^{-i\omega} (\eta_L+\eta_R)^{i\omega-\frac 12}  \, J_{i\kappa}(\eta_L+\eta_R) \ . \label{tilde-v}
\end{align}
An $\epsilon$-prescription that regularises the integral and the constant $E$ are given in Appendix \ref{A:Fourier transformations}. On the other hand, the modes $\hat v_{\omega,\kappa}$ found in \eqref{v-matrix-element} read
\begin{equation}\label{1-y}
\hat v_{\omega,\kappa} = B\, \eta_L^{\,i\tau-i\omega}\,\eta_R^{-i\tau}(\eta_L+\eta_R)^{i\omega-\frac 12}  \, J_{i\kappa}(\eta_L+\eta_R)\,,
\end{equation}
with $\, B = c_{\omega,\kappa}(-1)^{i\tau-i\omega} C/(2\pi)\,$. The two sets of solutions do not coincide, but of course they can be expanded in one another.

Another way of characterising these two sets of functions is by operators that they diagonalise. Both sets are eigenfunctions of the fuzzy Laplacian, which is in variables $\chi,\zeta$ given by
\begin{equation}\label{Fourier-Laplacian-reduced}
    \Delta = - \chi^2\partial_\chi^2 - \zeta^2\partial_\zeta^2 - 2\chi\zeta\partial_\chi\partial_\zeta - 2\, (\chi\partial_\chi + \zeta\partial_\zeta) - 4 \chi^2\ .
\end{equation}
The eigenvalue of both $w_{\omega,\kappa}$ and $\hat v_{\omega,\kappa}$ is, recall, $\kappa^2+1/4$. In addition, $w_{\omega,\kappa}$ are eigenfunctions of $\zeta\partial_\zeta$. On the other hand, $\hat v_{\omega,\kappa}$ are eigenfunctions of the operator
\begin{equation}\label{quantum-dilation}
    \hat D = (\chi+\zeta)\partial_\zeta - 2i\tau \frac{\chi}{\chi-\zeta} = \left(D^L - \frac12\right) - \frac{\eta_L}{\eta_R} \left(D^R-\frac12\right)\ .
\end{equation}
Recall that in the commutative case, the modes $\, v_{\omega,\kappa}$ are characterised as simultaneous eigenfunctions of $\{\Delta_{dS_2},D\}$. The fact that in the noncommutative case $\hat D$ differs from the sum $D^L + D^R$ is another manifestation of the fact that quantisation breaks the $SO(1,2)$ symmetry, discussed in the remark in Subsection \ref{ss3.1}. We expect that in higher dimensions, characterisation of noncommutative modes through differential equations similar to \eqref{quantum-dilation} is possible and can lead to considerable simplifications compared to analysis through integral representations.

\section{Summary and perspectives}
\label{S:Summary and perspectives}

In this work, we have performed a detailed study of the Laplacian on fuzzy de Sitter spaces and of its eigenfunctions, focusing on two and four dimensions. Our main results are 1) an explicit quantisation of solutions to the commutative Klein-Gordon equation and 2) a construction of a complete basis of solutions in the noncommutative setup.

The model of fuzzy de~Sitter space is based on the noncommutative frame formalism, which enables one to build geometry (connection, curvature, Laplacian) that resembles closely the commutative one. In addition, it uses symmetries of spacetime -- generators of the $SO(1,d)$ group, to define noncommutative coordinates and momenta as operators on the Hilbert space of a unitary irreducible representation of the group. Our quantisation of commutative solutions and the proof that it maps modes of the scalar field on commutative de~Sitter space to solutions of the fuzzy Klein-Gordon equation, rests on a particular choice of $(\upeta,\ty^i)$ coordinates. In fact, in cosmology, {\it physical} coordinates $\,\ty^i =a(\upeta) \tx^i= \tx^i/\upeta$ ($a(\upeta)$ denotes the scale factor) are used, along with {\it comoving} coordinates $\tx^i$, to calculate observational effects, but usually not in field quantisation. We have shown that solutions to the Klein-Gordon equation separated in $(\upeta,\ty^i)$ coordinates naturally lead to an $SO(1,d-1)$-invariant set of positive frequency modes and identified the choice of positive modes that corresponds to the Bunch-Davies vacuum.

In order to find all modes of the scalar field on the fuzzy de~Sitter space, we fixed the representation of the de~Sitter group\footnote{The representations that are used belong to unitary principal series and are labelled by $(\kappa,s)$. However, as the fuzzy Laplacian does not depend on spin $s$, the results/conclusions that we obtain are in fact generic and degeneracy in $s$ is easy to trace.}, realised operators as integral kernels, and solved the fuzzy Klein-Gordon equation as a differential equation in ordinary, commutative, variables. A remarkable fact is that, when written in a specific basis, this equation has the same structure as the commutative Klein-Gordon equation written in $(\upeta,\ty^i)$ coordinates. We did explicit calculations for heavy fields $M>(d-1)/2$, and they should be completed with a check of results for the light fields, as well as with details of the important case of a massless scalar field. 

The main feature that distinguishes between the fuzzy dS$_2$ and dS$_4$ is that in four dimensions the noncommutative space has many more degrees of freedom than its classical counterpart. While this property is manifest in our analysis, the questions of identifying ‘almost-classical' modes among the set of all fuzzy modes and the mechanism by which the higher KK-modes are suppressed remain highly non-trivial. We have made some initial progress on the first of these problems in the last section by characterising, in some cases, field modes through differential equations. The full resolution of these questions is our first task for the future. We believe they are amenable to weight-shifting techniques that have been extensively developed in conformal field theory, \cite{Costa:2011dw,Karateev:2017jgd,Buric:2022ucg}.

Our work is a first step towards formulation and analysis of quantum field theories on fuzzy de Sitter backgrounds. The simplest correlation function we may consider is the propagator. The field modes we obtained allow to write an integral representation for the latter. For example in two dimensions, we expand a quantum field as
\begin{equation}
    \Phi = \int d\omega \left(a_\omega \hat v_\omega + a^\dagger_\omega\hat v^\ast_\omega\right)\ .
\end{equation}
By promoting $a^\dagger_\omega$ and $a_\omega$ to creation and annihilation operators acting on a Fock space, in the same manner as in an ordinary quantum field theory on de~Sitter space, we arrive at the usual representation of the two-point function
\begin{align}\label{two-point-function}
    \langle \Phi_1 \Phi_2 \rangle & = \langle 0 | \int d\omega_1 \left( a_{\omega_1} \hat v_{\omega_1} + a^\dagger_{\omega_1}\hat v^\ast_{\omega_1}\right)  \int d\omega_2 \left( a_{\omega_2} \hat v_{\omega_2} + a^\dagger_{\omega_2}\hat v^\ast_{\omega_2}\right) |0\rangle\\
    & = \iint d\omega_1 d\omega_2\ \hat v_{\omega_1}\hat v_{\omega_2}^\ast \langle0|[a_{\omega_1},a_{\omega_2}^\dagger]|0\rangle = \int d\omega\ \hat v_\omega \hat v_\omega^\ast\ . \nonumber
\end{align}
Similarly as for the modes, we may encode the two-point function into its matrix elements. Working again in the conformal-time basis, this leads to a solvable integral, given in Appendix \ref{A:2pt}. An interesting question, that we leave for future work, is how this expression is related to the ‘quantised two-point function', which in the Bunch-Davies vacuum reads
\begin{equation}\label{BDfuzzy}
    \hat G_2 = \hat G_2(\heta,\hx,\heta',\hx') = \frac{\Gamma\left(\frac12+i\kappa\right)\Gamma\,\left(\frac12-i\kappa\right)}{4\pi}\ _2F_1\Big(\frac12+i\kappa,\frac12-i\kappa;1;\frac{1+\hat Z}{2}\Big)\ .
\end{equation}
The quantity $\hat Z$ in \eqref{BDfuzzy} is the quantum analogue of the geodesic distance between two points,
\begin{equation}
    Z(\xi,\xi') = \frac{\upeta^2 + \upeta'^2 - (\tx - \tx')^2}{2\upeta\upeta}\ .
\end{equation}
The corresponding operator is most naturally defined using symmetric ordering,\footnote{The primed and unprimed operators, corresponding to the first and the second point, commute with each other.}
\begin{align}
    \hat Z &= \frac12 \left(\heta\heta'^{-1} + \heta^{-1}\heta' + 2 \hy \hy' - \heta'^{-1} \{\heta^{-1},\hx^2\} - \heta^{-1} \{\heta'^{-1},\hx'^2\}\right)\\[2mm]
     & = \frac{q}{q'}\partial_q^2 + \frac{q'}{q}\partial_{q'}^2 + \left(\frac32-\mathbb{\Delta}\right)\left(\frac{1}{q'}\partial_q + \frac{1}{q}\partial_{q'}\right) + \frac{\frac12(q^2 + q'^2) + \mathbb{\Delta}(\mathbb{\Delta}-1) +\frac34}{qq'}\ .\nonumber
\end{align}

An important part of the analysis of the correlators in four dimensions, which would bring us to the initial goal of comparison with observations, is to work out the limiting behaviour of the two-point function at the reheating surface $\heta=0$.

In relation to this, we believe that the physical interpretation of correlation functions in QFTs on fuzzy de~Sitter spaces requires a further ingredient, that of semi-classical, coherent-like, states, \cite{Perelomov:1986tf}. In a similar context of fuzzy AdS$_3$, these states were defined in \cite{Buric:2022ton}, and the same construction applies here. Let us briefly describe it, focusing on the fuzzy dS$_2$. We start with some reference vector $|\xi_0\rangle\in\mathcal{H}$, with finite expectation values of coordinate operators,
\begin{equation}
    \upeta_0 = \langle\xi_0|\heta|\xi_0\rangle, \qquad \tx_0 = \langle\xi_0|\hx|\xi_0\rangle\,,
\end{equation}
and let $\xi_0 = (\upeta_0,\tx_0)$ be a point of the classical dS$_2$ manifold. Locally around $\xi_0$, any other point $\xi_1 = (\upeta_1,\tx_1)$ may be reached by a transformation of the form
 \begin{equation}
    \xi_1 = \lambda^{-e_0} e^{-b e_1} \xi_0 = \left(\lambda^{-1}\upeta_0,\tx_0 - b \upeta_0\right)\,,
\end{equation}
where, recall, $\{e_0,e_1\}$ is the classical frame. We define the semi-classical state ‘associated with $\xi_1$' by
\begin{equation}
    |\xi_1\rangle = \lambda^{\hp_0} e^{b \hp_1} |\xi_0\rangle\ .
\end{equation}
Local measurements at the point $\xi_1$ are to be modelled by expectation values of operators in the state $|\xi_1\rangle$. Axioms of the frame formalism, i.e. the frame relations, ensure that expectation values of Poincar\'e coordinate operators in the state $|\xi_1\rangle$ are equal to coordinates of the classical point $\xi_1$,
\begin{equation}\label{semi-classical-states-consistency}
    \langle\xi_1|\heta|\xi_1\rangle = \upeta_1, \qquad \langle\xi_1|\hx|\xi_1\rangle = \tx_1\ .
\end{equation}
Indeed, relations \eqref{semi-classical-states-consistency} are derived using the Baker-Campbell-Hausdorff formula, e.g.
\begin{equation*}
    \langle\xi_1|\heta|\xi_1\rangle = \langle\xi_0| e^{-b \hp_1}\, \lambda^{-\hp_0}\, \heta\, \lambda^{\hp_0}\, e^{b \hp_1}\, |\xi_0\rangle = \lambda^{-1} \langle\xi_0|e^{-b\hp_1}\, \heta\, e^{b\hp_1} |\xi_0\rangle = \lambda^{-1} \langle\xi_0|\heta|\xi_0 \rangle = \lambda^{-1}\upeta_0 = \upeta_1\ .
\end{equation*}
The same argument works for the coordinate $\hat x$. For observables other than Poincar\'e coordinates, analogues of \eqref{semi-classical-states-consistency} are not satisfied exactly but only approximately. In this way, the semi-classical states measure noncommutative corrections to any observable. In particular, the expectation values between pairs of semi-classical states, that is
\begin{equation}
    \langle \xi_1,\xi_2| \langle\Phi_1\Phi_2\rangle|\xi_1,\xi_2\rangle \,, \qquad \langle \xi_1,\xi_2|\hat G_2|\xi_1,\xi_2\rangle\,,
\end{equation}
will be the simplest examples of corrections that QFT observables receive due to noncommutativity. 

In the above discussion we restricted ourselves to fuzzy dS$_2$, for which the detailed computations will follow straightforwardly from results of the present work. Their extension to four dimensions, while proceeding along the same lines, is more challenging -- it is a major task for the future. A long term goal would be to include interactions and analyse three- or higher-point functions.

Finally, it would be worthwhile to carry the same analysis in three dimensions, for the fuzzy AdS$_3$ or the BTZ black hole of \cite{Buric:2022ton}. We have verified that all results from previous sections carry over, appropriately modified, to these spaces as well. Moreover, there are two considerable simplifications compared to fuzzy dS$_4$ -- coordinates $\hy^i$ actually commute, and the ‘internal space' is $S^1$ rather than $S^2$. This makes the fuzzy AdS$_3$ an ideal setup to study some of the quantum effects and to develop the remaining necessary methods for quantum field theory on fuzzy dS$_4$.

\paragraph{Acknowledgement} This work is funded by a research grant under the project H2020 ERC STG 2017 G.A. 758903 "CFT-MAP" and by 451-03-66/2024-03/200162 Development Grant of MNTRI, Serbia.

\appendix 

\section{Conventions for the de Sitter algebra}
\label{A:Conventions for the de Sitter algebra}

Here we give our conventions for the de Sitter group $G = SO(1,d)$ and its Lie algebra. The non-vanishing Lie brackets of $\mathfrak{g} = \mathfrak{so}(1,d)$ are
\begin{align}
    & [L_{ij},L_{kl}] = \delta_{jk} L_{il} - \delta_{ik} L_{jl} + \delta_{jl} L_{ki} - \delta_{il} L_{kj}\,,\\[2pt]
    & [L_{ij},P_k] = \delta_{jk} P_i - \delta_{ik} P_j, \quad [L_{ij},K_k] = \delta_{jk} K_i - \delta_{ik} K_j\,,\\[2pt]
    & [D,P_i] = P_i, \quad [D,K_i] = - K_i, \quad [K_i,P_j] = 2\left(L_{ij} - \delta_{ij}D\right)\ .
\end{align}
with $i,j=1,\dots,d-1$. The Lorentz-like generators $M_{\alpha\beta}$, $\alpha,\beta=0,\dots,d$ are related to the above by
\begin{equation}
    D = - i M_{0d}, \quad P_i = i (M_{0i} + M_{id}), \quad K_i = i (M_{0i} - M_{id}), \quad L_{ij} = i M_{ij}\ .
\end{equation}
They satisfy the bracket relations
\begin{equation}\label{brackets-so(1,d)}
    [M_{\alpha\beta},M_{\gamma\delta}] = i \left(\eta_{\alpha\gamma} M_{\beta\delta} - \eta_{\alpha\delta} M_{\beta\gamma} - \eta_{\beta\gamma} M_{\alpha\delta} + \eta_{\beta\delta} M_{\alpha\gamma}\right)\ .
\end{equation}
The quadratic Casimir and its value in the representation $(\mathbb{\Delta},l) = \pi_{\mathbb{\Delta},(l)}$ of $SO(1,d)$ are given by
\begin{equation}
    C_2 = -\frac12 M^{\alpha\beta} M_{\alpha\beta} = -D^2 -\frac12 \{P_i,K^i\} +\frac12 L^{ij} L_{ij}, \quad C_2(\mathbb{\Delta},l) = - \mathbb{\Delta}(\mathbb{\Delta}-d+1) - l(l+d-3)\ .
\end{equation}
In four dimensions, the de-Sitter group posses one more independent Casimir element, \eqref{quartic-Casimir-expression}. Its value in the representation $(\mathbb{\Delta},1/2)$, the only one that we make use of in this work, reads
\begin{equation}
    C_4\left(\mathbb{\Delta},1/2\right) = -\frac34 (\mathbb{\Delta}-1)(\mathbb{\Delta}-2)\ .
\end{equation}

\section{Details regarding kernels}
\label{A:Representation-theoretic background}

Here we give detailed definitions of kernels used in Section \ref{S:All fuzzy harmonics}. In particular, this explains the occurrence of minus sign in \eqref{v-matrix-element}. Consider first the case of $G=SO(1,2)$. Any element $\hat f$ of $\mathcal{A}$ is represented by a function $\hat f(q_L,q_R)$. The fact that the action of the Laplacian is written as \eqref{NC-Laplacian-general} means that the corresponding vector in $\mathcal{H}_{\mathbb{\Delta}} \otimes \mathcal{H}_{1-\mathbb{\Delta}}$ reads
\begin{equation}\label{expansion-in-tensor-product}
    \int dq_L dq_R\ \hat f(q_L,q_R) |q_L\rangle \otimes |q_R\rangle\,\in\mathcal{H}_{\mathbb{\Delta}} \otimes \mathcal{H}_{1-\mathbb{\Delta}}\ .
\end{equation}
Here, $|q_L\rangle\in\mathcal{H}_{\mathbb{\Delta}}$ and $|q_R\rangle\in\mathcal{H}_{1-\mathbb{\Delta}}$. On the other hand, regarding $\hat f$ as an operator $\mathcal{H}_{\mathbb{\Delta}} \to \mathcal{H}_{\mathbb{\Delta}}$, we can also expand it in terms of its matrix elements
\begin{equation}\label{expansion-in-matrix-elements}
    \hat f = \int dq_L dq_R\ F(q_L,q_R) |q_L\rangle \otimes \langle q_R|\,, \qquad F(q_L,q_R) = \langle q_L| \hat f| q_R\rangle\,,
\end{equation}
where $|q_L\rangle,|q_R\rangle\in\mathcal{H}_{\mathbb{\Delta}}$. We wish to relate $\hat f(q_L,q_R)$ and $F(q_L,q_R)$ in a way that respects the action of $SO(1,2)$. To this end, let $\varphi(q)\in\mathcal{H}_{\mathbb{\Delta}}$, i.e. the action on $P,D$ and $K$ on $\varphi$ is given by \eqref{conformal-generators-Fourier-1}-\eqref{conformal-generators-Fourier-2}. Then the action of $SO(1,2)$-generators on $\psi(q)$, defined by
\begin{equation}
    \psi(q) = \varphi(-q)^\ast\,,
\end{equation}
is \eqref{conformal-generators-Fourier-1}-\eqref{conformal-generators-Fourier-2} with $\mathbb{\Delta}=\frac 12 +i\tau $ replaced by $1-\mathbb{\Delta}=\frac 12 -i\tau$. In other words, $\psi(q)\in\mathcal{H}_{1-\mathbb{\Delta}}$. We deduce that $\hat f(q_L,q_R) = F(q_L,-q_R)$.
\smallskip

In four dimensions, there is an analogous statement that takes into account the spin degrees of freedom. Let $\mathcal{H}_{\mathbb{\Delta},(1/2)}$ be the carrier space of the representation \eqref{conformal-generators-Fourier-1}-\eqref{conformal-generators-Fourier-2} with $\Sigma$-matrices \eqref{4d-representation} and $\varphi = \varphi^a(q^i)$ an element of this space. Then, the function
\begin{equation}\label{sigma2-phi}
    \psi^a(q^i) = (\sigma_2)^a{}_b\, \varphi^b(-q^i)^\ast\,,
\end{equation}
belongs to $\mathcal{H}_{3-\mathbb{\Delta},(1/2)}\,$.

\section{Fourier transforms}
\label{A:Fourier transformations}

In this appendix we derive two formulas for the  Fourier transformations used in the main text. More precisely, we show how they can be obtained as regularisation of integrals given in the literature. The first one is used in \eqref{v-matrix-element}, 
\begin{equation}\label{Fourier-1}
    \int\limits_{-\infty}^{\infty } d\lambda\ e^{i\lambda q}\, (\lambda^2-1)^{-\frac{i\omega}{2}} \, Q^{-i\omega}_{-\frac 12+ i\kappa}(\lambda) \overset{\epsilon}{=} C \, q^{-\frac 12 +i\omega}\, J_{i\kappa}(q)\ .
\end{equation}
We start from the integral given in \cite{Watson},
\begin{equation} \label{Watson}
    \int\limits_0^\infty e^{-aq}\, J_{\tilde\nu}(bq) \, q^{\tilde\mu-1} \, dq = 2^{-\tilde \nu}\,
    \frac{\Gamma(\tilde\mu+\tilde\nu)}{\Gamma(\tilde\nu+1)}\ b^{\tilde\nu} \,a^{\tilde\mu -\tilde\nu-1}\, (a^2+b^2)^{\frac 12-\tilde\mu}\ 
    _2F_1\Big(\frac{\tilde\nu-\tilde\mu+1}{2}, \frac{\tilde\nu-\tilde\mu}{2}+1;\tilde\nu+1;-\frac{b^2}{a^2}  \Big)\,,
\end{equation}
which is valid when its parameters satisfy $\, {\rm Re}(\tilde\mu+\tilde\nu)>0$, $\, {\rm Re}\,a>0$, $\, {\rm Re}(a\pm ib)>0$. Choosing
\begin{equation}
    a=\epsilon +i\lambda, \quad b=1,\quad \tilde\nu =i\kappa,\quad \tilde\mu = \frac12 + i\omega,\quad \epsilon>0\,,
\end{equation}
we find that, apart from the integration boundaries, the integral gives the inverse Fourier transform of \eqref{Fourier-1} in the limit $\epsilon\to 0\,$. Formally, we can write
\begin{equation}
I(\lambda) =\lim_{\epsilon\to 0}  \,  \int\limits_0^\infty e^{-(\epsilon+i\lambda)q}\, J_{i\kappa}(q) \, q^{-\frac 12 +i\omega} \, dq   = \int\limits_0^\infty e^{-i\lambda q}\, J_{i\kappa}(q) \, q^{-\frac 12 +i\omega} \, dq  \ .
\end{equation}
Using a similar regularisation for the interval $(-\infty,0)$ and  property $\, J_{\tilde\nu}(-q) = e^{i\pi \tilde \nu} J_{\tilde\nu}(q) $, we find that 
$\      \int_{-\infty}^0 e^{-i\lambda q}\, J_{i\kappa}(q) \, q^{-\frac 12 +i\omega} \, dq = -e^{2\pi(\kappa + \omega)} \,I(\lambda) $. Therefore
\begin{equation}
    \int\limits_{-\infty}^\infty e^{-i\lambda q}\, J_{i\kappa}(q) \, q^{-\frac 12 +i\omega} \, dq =\left(1-e^{2\pi(\kappa + \omega)}\right)\, I(\lambda)\ .  
\end{equation}
The hypergeometric function in \eqref{Watson} can be expressed in terms of the Legendre function $Q\,$: using this relation, we obtain
\begin{equation}
I(\lambda)= \sqrt{2\pi}\,e^{-\frac{i\pi}{2}\, (\frac 12-i\omega + i\kappa)}\, 
    \frac{\Gamma\left( \frac 12+i\omega+ i\kappa\right)}{\Gamma\left( \frac 12-i\omega + i\kappa\right)}\,
    (\lambda^2-1)^{-\frac{i\omega}{2}}\, Q^{-i\omega}_{-\frac 12+i\kappa}(\lambda) \ ,
\end{equation}
and, consequently, \eqref{Fourier-1}. The last formula also determines the constant $C$,
\begin{equation}
  C  = \sqrt{2\pi}\ \frac{\Gamma\left( \frac 12-i\omega+ i\kappa\right)}{\Gamma\left( \frac 12+i\omega + i\kappa\right)} \frac{e^{\frac{i\pi}{2}(\frac 12-i\omega + i\kappa)}}{1-e^{2\pi(\kappa+\omega)}}\ .
\end{equation}

Another transformation that we define by regularisation is used to obtain \eqref{tilde-v}. It reads
\begin{equation}\label{Fourier-2}
     \int_{-\infty}^\infty \eta^{i\omega-1}\, e^{2i\zeta\eta} \,d \eta 
     \overset{\varepsilon}{=}
     - 2 \Gamma(i\omega)
      \, e^{\frac{\omega\pi}{2}}\, \sinh(\pi\omega)\, (2\zeta)^{-i\omega}
      \ .
\end{equation}
To derive it, we start from the formula, \cite{Erdelyi} 
\begin{equation}
    \int_0^\infty t^{\gamma-1}\, e^{-ct\cos \beta-ict\sin\beta} dt = \Gamma(\gamma) \,c^{-\gamma}\, e^{-i\gamma\beta}\,,
\end{equation}
which holds when
\begin{equation}
    -\frac \pi 2<\beta<\frac\pi 2,\quad {\rm Re}\,\gamma >0\, ,  \qquad {\rm or}\quad 
    \beta =\pm \frac \pi 2 ,\quad 0<{\rm Re}\,\gamma<1\ .
\end{equation}
We take \ $\gamma=\epsilon+i\omega $, $\ \beta=- \pi/ 2$, $\ c=2\zeta$, and define
\begin{equation}
\int\limits_0^\infty \eta^{i\omega-1}\, e^{2i\zeta \eta} \, d\eta =\lim_{\epsilon\to 0}  \,  \int\limits_0^\infty \eta^{\epsilon+i\omega-1}\, e^{2i\zeta \eta} \, d\eta  = \Gamma(i\omega) \,(2\zeta)^{-i\omega}\, e^{-\frac{\omega\pi}{2}}\ .
\end{equation}
Similarly, $\ \int_{-\infty}^0 \eta^{i\omega-1}\, e^{2i\zeta\eta} \,d \eta= (-1)^{i\omega-1} \Gamma(i\omega) \,(2\zeta)^{-i\omega}\, e^{\frac{\omega\pi}{2}}\ $, and we obtain \eqref{Fourier-2}. Combining \eqref{tilde-v} with \eqref{Fourier-2} gives the value of the remaining constant $E$,
\begin{equation}
    E= -4 (-1)^{-i\omega} \, e^{\frac{\pi\omega}{2}} \sinh(\pi\omega) \, \Gamma(i\omega) \,C\ .
\end{equation}

\section{Special functions}
\label{A:Special functions}

In this appendix, we list some properties of the special functions used throughout the work. The Bessel functions of the first kind $J_{\pm i\kappa}(z)$ and the Hankel functions $H^{(1)}_{i\kappa}(z)$ i $H^{(2)}_{i\kappa}(z)$ are related as
\begin{equation}
    H^{(1)}_{i\kappa}(z)=\frac{e^{\pi \kappa} J_{i\kappa}(z)-J_{-i\kappa}(z)}{\sinh{(\pi \kappa)}}\, ,  \qquad H^{(2)}_{i\kappa}(z)=\frac{J_{-i\kappa}(z)-e^{-\pi \kappa}J_{i\kappa}(z)}{\sinh{(\pi \kappa)}} \ .
\end{equation}
Bessel functions satisfy  orthogonality relation
\begin{equation}\label{Bessel-normalisation}
    \int\limits_0^{\infty} dx\, x\, J_{\nu}(kx) J_{\nu}(k'x) = \frac1k \delta(k-k'), \quad \nu>-\frac12\,,
\end{equation}
and, asymptotically for $z\to0$, behave as $\ J_{\nu}(z)\sim (\frac{z}{2})^\nu /\Gamma(1+\nu)\, $, $ \ \nu \not=-1, -2, -3, \dots$ .

Legendre functions of the second kind may be defined in terms of the hypergeometric function,
\begin{equation}\label{Legendre-function-definition}
  Q^\mu_\nu(z) =  \sqrt{\pi}\, \frac{ e^{i\mu\pi}}{2^{\nu+1}}\,\frac{\Gamma(\mu+\nu+1)}{\Gamma(\nu+\frac 32)}\,
  z^{-\nu-\mu-1} (z^2-1)^{\frac \mu 2}\,\, _2F_1\Big( \frac{2+\nu+\mu}{2}, \frac{1+\nu+\mu}{2} ;\nu+\frac 32;\frac{1}{z^2}  \Big)\, ,\quad |z|>1\ .
\end{equation}
In Wolfram Mathematica, this definition is implemented as LegendreQ$[\nu,\mu,3,z]$, the third argument standing for `type three'. It distinguishes \eqref{Legendre-function-definition} from other commonly used definitions.

For spherical harmonics we use the quantum mechanical conventions,
\begin{equation}
    Y_l^m(\vartheta,\phi) = (-1)^m \sqrt{\frac{(2l+1)(l-m)!}{4\pi(l+m)!}} e^{i m \phi} P^m_l(\cos\vartheta)\,,
\end{equation}
which ensure the following  properties
\begin{equation}
  (Y_l^m)^\ast = (-1)^m Y_l^{-m}\ , \qquad  Y_l^m(\pi-\vartheta,\pi+\phi) = (-1)^l Y_l^m(\vartheta,\phi)\ .
\end{equation}
Useful relations are the addition theorem and the completeness relation,
\begin{align}\label{addition-thm-spherical-harmonics}
 &\sum_{m=-l}^l Y_l^m(\vartheta_L,\phi_L) \, Y_l^m(\vartheta_R,\phi_R)^\ast = \frac{2l+1}{4\pi}\, P_l(\cos\Theta)\,,
\\[4pt]
  & \sum_{l,m} Y_l^m(\vartheta_L,\phi_L) \, Y_l^m(\vartheta_R,\phi_R)^\ast = \frac{1}{\sin\vartheta_L}\, \delta(\vartheta_L - \vartheta_R) \,\delta(\phi_L - \phi_R)\,,
\end{align}
with
\begin{equation}
    \cos \Theta \equiv \Omega = \frac{\vec{q}_L \cdot \vec{q}_R}{q_L q_R} =\cos\vartheta_L \cos\vartheta_R + \cos(\phi_L - \phi_R)\sin\vartheta_L\sin\vartheta_R\ .
\end{equation}
In conventions for spinor harmonics we follow \cite{Bjorken:100769},
\begin{equation}
    \varphi_{j,m}^{+}(\vartheta,\phi) = \begin{pmatrix}
        \sqrt{\frac{j+m}{2j}} Y_{j-1/2}^{m-1/2}(\vartheta,\phi)\\[8pt]
        \sqrt{\frac{j-m}{2j}} Y_{j-1/2}^{m+1/2}(\vartheta,\phi)
    \end{pmatrix}, \qquad \varphi_{j,m}^{-}(\vartheta,\phi) = \begin{pmatrix}
        \sqrt{\frac{j+1-m}{2(j+1)}} Y_{j+1/2}^{m-1/2}(\vartheta,\phi)\\[8pt]
        -\sqrt{\frac{j+1+m}{2(j+1)}} Y_{j+1/2}^{m+1/2}(\vartheta,\phi)
    \end{pmatrix}\ ,
\end{equation}
Here, $ j=\frac12,\frac32, \dots$ and $m=-j,\dots ,j\,$. Spinor harmonics satisfy
\begin{equation}
    \varphi_{j,m}^{\pm} = \frac{\sigma^i q_i}{q}\, \varphi_{j,m}^{\mp}\,, \qquad \varphi_{j,m}^{\pm}(\pi-\vartheta,\pi+\phi) = (-1)^{j\mp\frac12}\, \varphi_{j,m}^{\pm}(\vartheta,\phi)\ .
\end{equation}
Defining functions $\psi^\mathrm{a}_{j,m}$ in accordance with \eqref{sigma2-phi}, we find
\begin{equation}
    \psi^\mathrm{a}_{j,m}(\vartheta,\phi) \equiv \sigma_2 \,\varphi^{\mathrm{a}}_{j,m}(\pi-\vartheta,\pi+\phi)^* = i\, (-1)^{j+m+1}\, \varphi^{\mathrm{a}}_{j,-m}(\vartheta,\phi) \ .
\end{equation}
Spinor harmonics satisfy various addition formulas, including
\begin{align}
  &  \sum_m\varphi^\uparrow_{j,m}(\vartheta_L,\phi_L) (\sigma_2 \varphi_{j,m}^\uparrow(\vartheta_R,\phi_R))^\ast = T\,\frac{i(2j+1)}{8\pi} \left(P_{j+\frac12}(\Omega) + P_{j-\frac12}(\Omega)\right)  \begin{pmatrix}
       \tan \frac \Theta 2 & -1\label{add-formula-1}\\[4pt]
        0& 0
    \end{pmatrix}\,,\\[12pt]
  & \sum_m\varphi^\downarrow_{j,m}(\vartheta_L,\phi_L) (\sigma_2 \varphi_{j,m}^\downarrow(\vartheta_R,\phi_R))^\ast = T\,\frac{i(2j+1)}{8\pi} \left(P_{j+\frac12}(\Omega) + P_{j-\frac12}(\Omega)\right)  \begin{pmatrix}
    0 & 0\\[4pt]
    1&    \tan\frac\Theta 2
    \end{pmatrix}\ .\label{add-formula-2}
\end{align}
Here, $T$ is the matrix such that the most general rotation invariant kernel $f^a{}_b(\vec{q}_L,\vec{q}_R)$, i.e. satisfying $(L_{ij}^L + L_{ij}^R) \hat f(\vec{q}_L,\vec{q}_R)^a{}_b = 0$ takes the form
\begin{equation}\label{general-solution-Ward-identities}
    f^a{}_b(\vec{q}_L,\vec{q}_R) = (T_1)^a{}_c (T_2)_b{}^d g^c{}_d(q_L,q_R,\Omega) \equiv T^{ad}{}_{cb} g^c{}_d(q_L,q_R,\Omega)\,,
\end{equation}
for arbitrary functions $g^c{}_d$. The addition formulas \eqref{add-formula-1}-\eqref{add-formula-2} are useful in passing from matrix elements of rotation-invariant kernels in the time operator eigenbasis used in the main text, to functions $f(\vec{q}_L,\vec{q}_R)^a{}_b$. For completeness, we spell out the explicit expression for $T_1$ and $T_2$
\begin{equation}
    T_1 = \begin{pmatrix}
        1 & \frac{e^{i\phi_R} \sin\Theta \sin\vartheta_L}{X}\\
        e^{i\phi_L}\tan\frac{\vartheta_L}{2} & \frac{-2 e^{i(\phi_L+\phi_R)}\sin\Theta\cos^2\frac{\vartheta_L}{2}}{X}
    \end{pmatrix}\,, \quad T_2 = \begin{pmatrix}
        \frac{-e^{-i(\phi_L+\phi_R)}X}{2\sin\Theta} & -\frac12 e^{-i\phi_L} \sin\vartheta_L\\
        \frac{-e^{i\phi_R}X\tan\frac{\vartheta_L}{2}}{2\sin\Theta} & \cos^2 \frac{\vartheta_L}{2}
    \end{pmatrix}\,,
\end{equation}
where
\begin{equation}
    X = e^{i\phi_R} (\Omega + \cos\vartheta_R)\sin\vartheta_L - 2 e^{i\phi_L}\cos^2\frac{\vartheta_L}{2}\sin\vartheta_R\ .
\end{equation}

\section{Two-point function integral}
\label{A:2pt}

Here we compute the integral giving the two-point function on the fuzzy dS$_2$, \eqref{two-point-function}. The two-point function in the $q$-basis is
\begin{align}
    & \langle q_L^{(1)}, q_L^{(2)} |\langle\Phi_1 \Phi_2\rangle|q_R^{(1)}, q_R^{(2)}\rangle = \int d\omega\ v_\omega\left(q_L^{(1)},q_R^{(1)}\right) v^\ast_\omega\left(q_L^{(2)},q_R^{(2)}\right)\\
    & = \frac{1}{4\pi} \left(\frac{q_L^{(1)} q_R^{(2)}}{q_L^{(2)} q_R^{(1)}}\right)^{i\tau} \frac{J_{i\kappa}(q_L^{(1)}+q_R^{(1)}) J^\ast_{i\kappa}(q_L^{(2)}+q_R^{(2)})}{\sqrt{(q_L^{(1)}+q_R^{(1)})(q_L^{(2)}+q_R^{(2)})}} \int d\omega\ |c_{\omega,\kappa}|^2 |C|^2 \left(\frac{q_L^{(2)} (q_L^{(1)} + q_R^{(1)})}{q_L^{(1)} (q_L^{(2)} + q_R^{(2)})}\right)^{i\omega}\ . \nonumber
\end{align}
We focus on the last integral over $\omega$, denoting it $\mathcal{I}(Q)$, where
\begin{equation}
     e^{2\pi Q} = \frac{q_L^{(2)} (q_L^{(1)} + q_R^{(1)})}{q_L^{(1)} (q_L^{(2)} + q_R^{(2)})}\ .
\end{equation}
By substituting for constants $C$ and $c_{\omega,\kappa}$, one obtains
\begin{align}
    \mathcal{I}(Q) & = \frac{1}{8\pi\sinh(\pi\kappa)} \int d\omega\ \frac{\cosh(\pi(\omega+\kappa))}{\sinh^2(\pi(\omega+\kappa))} e^{-3\pi(\omega+\kappa)}e^{2i\pi Q\omega}\\
    & = \frac{e^{-2i\pi\kappa Q}}{\pi} \left((3-2i Q)B_{e^{2\pi(\omega+\kappa)}}(-1+iQ,0) - \frac{e^{\pi(2iQ-3)(\omega+\kappa)}}{\sinh(\pi(\omega+\kappa))}\right)\Big|_{\omega_1}^{\omega_2}\ . \nonumber
\end{align}
After the appropriate regularisation, the integral gives the two-point function.

\end{document}